\documentclass[twocolumn]{aastex631}
\usepackage{booktabs}
\usepackage{longtable}
\usepackage{tikz}
\usepackage{amsmath}
\DeclareSymbolFont{matha}{OML}{txm i}{m}{it}% txfonts
\DeclareMathSymbol{\varv}{\mathord}{matha}{118}
\usepackage{graphicx}
\usepackage{appendix}
	
\usepackage{xcolor}

\newcommand{\Harvard}{Center for Astrophysics \textbar{} Harvard \& Smithsonian, 60 Garden Street, Cambridge, MA 02138-1516, USA}
\newcommand{\IAIFI}{The NSF AI Institute for Artificial Intelligence and Fundamental Interactions}
\newcommand{\Purdue}{Department of Physics and Astronomy, Purdue University, 525 Northwestern Avenue, West Lafayette, IN 47907-2036, USA}
\newcommand{\UCSC}{Department of Statistics, University of California, Santa Cruz}

\newcommand{\NI}{$\rm ^{56}Ni$}

\newcommand{\he}{$\rm ^{4}He$}
\newcommand{\mni}{$m_{\rm Ni}$}
\newcommand{\mhe}{$m_{\rm He}$}
\newcommand{\mop}{$m_{\rm bulk}$}

\newcommand{\msun}{$\rm M_{\odot}$}

\newcommand{\mej}{$m_{ej}$}
\newcommand{\vej}{$v_{ej}$}

\newcommand{\sed}{$\texttt{SEDONA}$}
\newcommand{\cmf}{$\texttt{CMFGEN}$}
\newcommand{\mos}{$\texttt{MOSFiT}$}
\newcommand{\Dp}{$\rm \eta_{vel}$}

\newcommand{\Rhp}{$\rm \eta_{He}$}
\newcommand{\Rnp}{$\rm \eta_{Ni}$}
\newcommand{\Ropp}{$\rm \eta_{bulk}$}

\newcommand{\gamn}{$r_{\rm 95}$}
\newcommand{\gamnh}{$r_{\rm 95, He}$}
\newcommand{\gamnn}{$r_{\rm 95, Ni}$}
\newcommand{\gamnop}{$r_{\rm 95, bulk}$}

\newcommand{\lc}{light curve}

\newcommand{\lcs}{light curves}
\newcommand{\opacity}{bulk}

\begin{document}

\title{Radiative Transfer Modeling of Stripped-envelope Supernovae II: Neural Network Emulation of Light Curves}

\author[0000-0002-0840-6940]{S.~Karthik~Yadavalli}
\affiliation{\Harvard}

\author[0000-0002-5814-4061]{V.~Ashley~Villar}
\affiliation{\Harvard}
\affiliation{\IAIFI}

\author[0000-0001-7081-0082]{Maria R. Drout}\affiliation{David A. Dunlap Department of Astronomy and Astrophysics, University of Toronto, 50 St. George Street, Toronto, Ontario, M5S 3H4, Canada}

\author[0000-0001-6395-6702]{Sebastian Gomez}
\affiliation{Department of Astronomy, The University of Texas at Austin, 2515 Speedway, Stop C1400, Austin, TX 78712, USA}

\author[0000-0001-7676-948X]{Miranda Pikus}
\affiliation{\Purdue}

\author[0000-0003-2779-6507]{Yunyi Shen}
\affiliation{\UCSC}

\begin{abstract}
We present the first neural-network emulator of stripped-envelope supernova (SESN) \lcs, trained on a grid of 4499 light curves simulated with the radiative transfer (RT) code \sed. Using this emulator, we show that \mni, \mej, the ejecta velocity profile, and the degree of \NI~mixing can all be inferred from multiband \lcs. We find that the degeneracy between ejecta mass and ejecta velocity is substantially weaker with this emulator than in traditional semianalytical models. The emulator is able to independently constrain the influence of ejecta mass and of ejecta velocity on the resulting \lc, rather than making them degenerate by design as traditional semianalytical models do. We additionally show that this inference is significantly more accurate than that done by the classical Arnett model for both simulated ZTF-like and LSST-like \lcs. Finally, we present \lc~fits to three well-studied SESNe: SN~1994I, SN~2007gr, and iPTF13bvn, constraining their \mni, \mej, and \NI~mixing.

\end{abstract}


\section{Introduction}
\label{sec:intro}

Stripped-envelope supernovae (SESNe) are stellar explosions that arise from massive stars whose outer envelope has been at least partially stripped prior to explosion (see e.g., \citealt{Smartt2009} for a review). The mechanism of this outer envelope stripping and therefore the progenitor system of hydrogen-free SESNe (specifically Type Ib and Type Ic supernovae) has been debated in literature \citep{Wheeler1985, Gaskell1986, Begelman1986, Woosley1993, nomoto1995evolution, podsiadlowski1992presupernova, Woosley1995, kobulnicky2007new}. Though wind-driven mass loss from massive Wolf-Rayet (WR) stars was the originally proposed stripping mechanism \citep{Wheeler1985, Gaskell1986, Begelman1986}, a binary progenitor star, where the exploding star's outer layers are stripped by the companion star, is likely a more prevalent progenitor \citep{nomoto1995evolution, podsiadlowski1992presupernova, Woosley1995, kobulnicky2007new}.  

SESN \lcs~span a relatively large range of luminosities, evolution timescales, and color \citep{Drout2011, Taddia2015, Lyman2016, Taddia2018}. Explaining this full range only using \NI~decay would require unphysical \NI~and ejecta masses, suggesting possible SESN power sources beyond the canonical radioactive decay of \NI~\citep{ertl2020, Sharon2020, afsariardchi2021, Rodriguez2024}. Regardless of the underlying heating source, the optical light curve traces thermalized radiation diffusing through dense, fast-moving, and optically thick ejecta \citep{Colgate1969, Arnett1982}. The radiation detected from these events can reveal much about the composition of the SESN ejecta, the energetics of the explosion, and, potentially, fundamental properties of the progenitor system (see e.g. \citealt{Arnett1982, Khatami2019, afsariardchi2021}).

Due to several dedicated transient surveys over the past $\sim$decade (such as ATLAS, PAN-STARRS, the Young Supernova Experiment, and the Zwicky Transient Facility; \citealt{Kaiser2002, Atlas_cite, masci2019zwicky, yse_ref_3, aleo2023young}), hundreds of multiband stripped-envelope supernova (SESN) \lcs~have now been observed with sufficient light curve coverage for physical inference. With the advent of the Legacy Survey of Space and Time (LSST) conducted by the Vera C. Rubin Observatory, $\sim$60k SESNe are expected to be discovered every year \citep{Kessler2019}.

The mapping from SESN progenitor characteristic to the observed \lc~is challenging. Traditionally, semianalytical models like the so-called ``Arnett model'' \citep{Arnett1982, Chatzopoulos2012} are used. However, currently such models lack many known complexities, such as frequency dependent opacities and variable ejecta profiles (e.g., \NI~mixing; see \citealt{dessart2012, dessart2015, Khatami2019, afsariardchi2021}). Importantly, the systematic errors in the inferred physical parameters caused by these simplified assumptions are not fully understood. \cite{Khatami2019} and \cite{Yadavalli2026} (hereafter Y26) both attempt to characterize the uncertainty/bias in the \mni~inferred by the Arnett model. \cite{Khatami2019} account for these effects by introducing a dimensionless parameter $\beta$ that absorbs the light-curve physics neglected by the Arnett model.%: the spatial distribution of radioactive heating (e.g., \NI~mixing) and the effects of recombination on the ejecta opacity.

More realistic SESN \lcs~can be simulated using radiative transfer (RT) integration codes such as \texttt{STELLA} \citep{Blinnikov1993, Blinnikov2006}, \texttt{SNEC} \citep{SNEC_exp2015}, \texttt{sedona} \citep{Kasen2006}, and CMFGEN \citep{CMFGEN_description}. These codes track how photons diffuse through ejecta described with a prescription of velocity, density, and chemical abundances. As such, RT simulation codes can be used to more sensitively probe the impact SESN ejecta physics has on the resulting SESN \lc~(see e.g. \citealt{dessart2012, dessart2015, dessart2016, Harvey2025}). Y26 present a grid of SESN \lcs~and explore the relationships between physical parameters and resulting \lcs. They find hints that SESN \lcs~are shaped not just by the integrated \mej~and \mni~but also by how this mass is \textit{distributed} throughout the ejecta; this suggests that these properties can potentially be inferred using RT simulation-based \lc~modeling. They find that \NI~``mixing'' is an interplay between how much \NI~is mixed into outer ejecta layers and how steeply the velocity profile varies throughout the ejecta layers. However, it is unclear if such physical inferences are recovered even with rigorous Bayesian fitting techniques for realistic SESN \lcs. 

However, RT simulations are expensive. An inference pipeline built on RT simulations to compare models to observed SESN \lcs~is computationally unfeasible unless the cost of the call to the forward model is reduced from $\sim$hours to $\sim$milliseconds. In order to reliably sample posterior distributions on physical parameters using a Bayesian inference framework, it may be necessary to perform up $\sim1$M calls to the forward model. Inference using current RT simulation codes would take in excess of $2000$~years on a single core. The call to the forward model can be dramatically sped up using emulators, or computationally cheap approximators, of RT simulation codes.

In recent years, neural network-based emulators have been used to approximate the output of RT simulation codes. In particular, several emulators of the RT code \texttt{tardis} have been presented in literature to perform Bayesian inference on observed spectra of SNe \citep{Kerzendorf2021, OBrien2024, lu2026traces}. \cite{OBrien2024} probe the ejecta properties of Type Ia SNe, and they find that 1991T-like SNe~Ia may be otherwise typical SNe~Ia with an underabundance of intermediate mass elements in their outer ejecta. \cite{lu2026traces} construct an emulator for SESNe spectra to perform abundance and density profile measurements of the Type Ic SN~2014L. With this emulator, they detect trace signatures of helium. Similarly, \cite{peng2024kilonova} construct an emulator for the \lcs~of kilonovae --- the thermalized transient following a neutron star merger --- trained on simulations from the RT code \texttt{supernu} \citep{Wollaeger2014}. With it, they fit the \lc~of AT2017gfo via a model with a lanthanide-rich, equatorial dynamical component and a lanthanide-poor, polar wind component viewed at some inclination. %They find that the \lc~robustly constrains the mass and velocity of the dynamical (equatorial) ejecta, but only weakly constrains the wind (polar) component and the viewing angle, indicating that the \lc~alone poorly determines the polar geometry of the ejecta. 
Although their emulator reproduces the observed \lc~well, the physical parameters they recover differ from previous studies that rely on semianalytic formalisms, specifically requiring a notably lower ejecta mass. Crucially, they also find that two emulators trained on the same library of \lcs~find two distinct fits to the observed \lc. They observe that two emulators, even when trained on the same underlying dataset, can settle in different final solutions because their training process started from different initial conditions. They claim that this is strong motivation for accurately characterizing the systematic error in emulator fits. 

Here, we present the first neural network-based emulator for SESN light curves, trained on the broad set presented in Y26. In Section~\ref{sec:methods}, we briefly review the RT implementation and the set up of the physical parameters explored by grid of SESN \lcs. We then detail the architecture, training process, and evaluation of the emulator over the grid of \lcs. In Section~\ref{sec:fitting_sim}, we present fits to simulated \lcs~with realistic observational cadence and uncertainties to characterize the fidelity of physical inference using the emulator. In Section~\ref{sec:fitting_real}, we present fits to three observed SESN \lcs. Finally we enumerate our conclusions in Section~\ref{sec:conclusion}.

\section{Emulator Construction and Evaluation}
\label{sec:methods}
Here we detail the construction of the SESN \lc~emulator which we present in this paper. All synthetic (i.e. either simulated or emulated) \lcs~in this work are powered only by \NI~decay; no other power sources (such as central engines or shocks) are considered in calculating SESN \lcs. In addition, local thermal equilibrium is assumed at all epochs in all simulations presented here (see Section~\ref{subsec:sed_uncertainties}). 

\subsection{Training the Emulator}
Y26 construct a novel physical parameterization for SESN ejecta profiles. Building on the works of \cite{woosley2019, ertl2020, Woosley2021} who evolve, explode, and simulate the SNe resulting from 81 helium stars, Y26 show that a SESN ejecta profile can be parameterized uniquely using nine numbers. The nine parameters used to parameterize SESN ejecta profiles are \mhe (mass of \he~in the ejecta), \mni~(the mass of \NI~ in the ejecta), \mop~(the mass of the ejecta excluding \he~and \NI), inner ejecta velocity ($\rm vel_{min}$), difference between outer and inner ejecta velocity ($\rm \Delta_{vel}$), and four other numbers (\Dp, \Rhp, \Rnp, and \Ropp) that describe the how each of these values are \textit{distributed} in the ejecta. The values \Dp, \Rhp, \Rnp, and \Ropp~are latent variables that, when passed into an autoencoder, produce the ejecta distributions of velocity, \he~mass, \NI~mass, and \opacity, respectively.  Y26 detail how each of these physical parameters influence the resulting \lc. For example, they find mixing either of \NI~or the \opacity~mass into the higher-velocity ejecta layers causes the \lc~to redden more quickly. Y26 also present a grid of 1000 \lcs~constructed by randomly and independently sampling each of the nine physical parameters from a uniform prior distribution. The prior distribution (detailed in Table 1 of Y26) for each parameter captures the full range of that parameter from the 81 \cite{Woosley2021} models. 
 
 Because the values \Rhp, \Rnp, \Ropp, and \Dp~are fed into a neural network to produce meaningful ejecta profiles, how each of these parameters impacts the resulting ejecta distribution is not easily interpretable. Y26 present several summary statistics that make it possible to better interpret the different mass distributions.  These are \gamnh, \gamnn, and \gamnop: the ratio of the ejecta velocity containing $95\%$ of the \he, \NI, and \opacity~mass to the maximum ejecta velocity, respectively. An ejecta profile with, for eample, \NI~that is considerably mixed outwards will have a larger \gamnn.  In this paper, we often use \gamn~instead of $\rm \eta$ to describe the mass distributions.

In this work, we construct a neural network-based emulator over the Y26 range of SESN models. The emulator is trained over a set of 4499 models simulated in \sed, augmenting the original Y26 grid of $1000$ \lcs~by drawing $3499$ additional points from the same prior space. Because \sed~runs earlier than $5$ days after explosion are prohibitively expensive, each calculation starts at 5 days and runs until 60 days after explosion with a timestep of 0.2 days \footnote{Calculating starting from just 4 days after explosion instead of 5 days already almost doubles the simulation time for most models. While it is true that much physical information is revealed in the \lc~rise, most SESNe in our grid take 15-20 days to reach peak after explosion. As such, we find that starting from day 5 is sufficient to constrain physics.}. At each timestep, the spectrum is evaluated over a grid of 602, log-uniformly spaced, wavelengths (corresponding to R $\approx$500 at $\rm 5000 \AA$), from $3000$\AA~to $10000$\AA. The network is trained to predict the spectrum produced by a given ejecta profile at an inputted time. We ultimately train the network to predict a downsampled version of the spectrum (R$\approx30$) due to substantial noise when predicting the full-resolution simulations. These downsampled spectra have only 41 log-uniformly spaced wavelengths over the same wavelength range. Training on downsampled spectra does not result in a loss in accuracy of the emulated \lc.  The network takes ten inputs: the nine physical parameters described above and time after explosion.

The emulator has a multilayer perceptron (MLP) architecture, with five hidden layers and 2000 neurons per layer. The network was trained with a learning decay rate of 0.85, batch size of 512, and for 150 epochs. The number of hidden layers, neurons per layer, learning decay rate, batch size and number of epochs are hyperparameters of this training process. To find the optimal combination of these hyperparameters we perform a grid search over these six dimensions. We find that the performance of the emulator is only loosely dependent on the architecture of the network and that the emulator performance is instead largely limited by the grid of \sed~models. Table~\ref{table:hyperparameters} shows the combinations of parameters we search to find the optimal network.

\begin{table}
    \centering
    \caption{Hyperparameter grid search values for the MLP emulator. 
    The chosen value for each hyperparameter is shown in the rightmost 
    column.}
    \label{table:hyperparameters}
    \begin{tabular}{lcc}
        \toprule
        Hyperparameter & Values Searched & Chosen Value \\
        \midrule
        Hidden Layers         & 3, 4, 5, 6, 7                  & 5     \\
        Neurons per Layer     & 500, 1000, 1500                & 2000  \\
                              & 2000, 2500                     &       \\
        Learning Decay Rate   & 0.01, 0.1                      & 0.85  \\
                              & 0.5, 0.7, 0.85                 &       \\
        Batch Size            & 256, 512, 1024                 & 512   \\
        Epochs                & 50, 100, 150                   & 150   \\
                              & 200, 250, 300                  &       \\
        \bottomrule
    \end{tabular}
\end{table}

Training the emulator is a multi-step process, where we use two approaches to augment the training set of \sed~simulations: 1) uniformly randomly sampling and simulating new points in the space and 2) active learning. In active learning, we iteratively add newly simulated models to the training set and incrementally increase the emulator performance. We first train and evaluate the emulator on an 80/20 train/test split on the original grid of Y26 1000 \lcs. We then identify the ten simulations with the worst emulator reconstruction. For each of these ten samples, we randomly resample 30 new points in its neighborhood. We sample new points around each point by drawing from a nine-dimensional Gaussian, whose standard deviation in each dimension is equal to $1\%$ of the range of that dimension. As such, each iteration adds 300 newly simulated models to the training set. Then, a new iteration of the emulator is trained on a new 80/20 train/test split of the whole grid, and the next iteration of active learning ensues. We find that while our implementation of active learning is good for improving emulator performance in a specific region of parameter space, it largely does not improve the overall emulator performance throughout the space. We rely on random sampling more to augment the training set than on active learning. In the end, we augmented our original training set with $2979$ models from random sampling and  $520$ models from active learning.

We note here that we also experimented with a custom transformer neural network-based emulator. We found that the transformer performed remarkably well at reconstructing spectra, even without downsampling. However, we ultimately opted for the MLP architecture because forward model calls to MLPs are much faster ($\sim$~milliseconds) than transformer model calls ($\sim100$s of milliseconds).

\subsection{Emulator Performance and Systematic Uncertainties}
\label{subsec:sed_uncertainties}
\begin{figure}
    \centering
    \includegraphics[width=\linewidth]{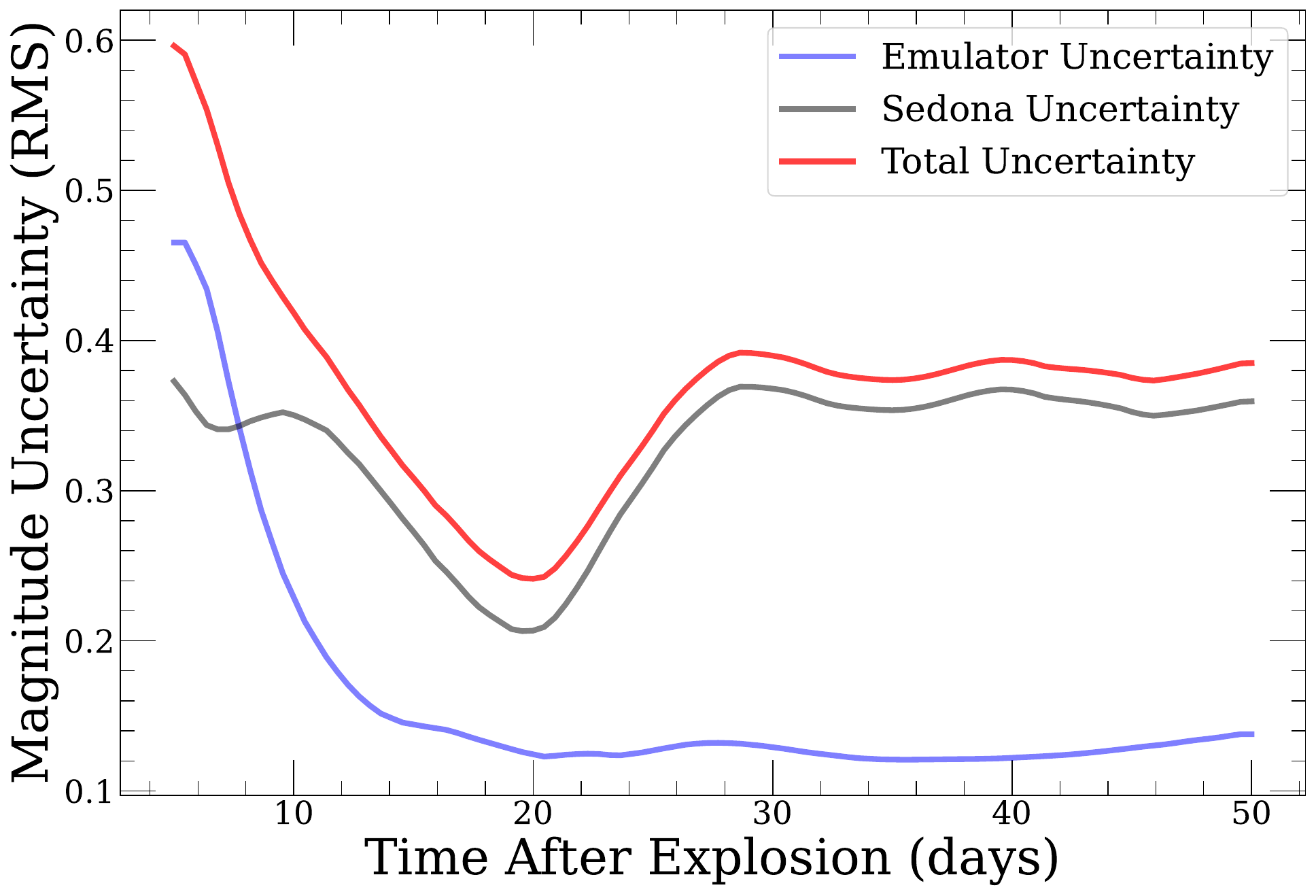}
    \caption{Systematic uncertainty of the inference pipeline as a function of time since explosion. The emulator uncertainty (blue), measured as root mean square of difference between predicted and true magnitude averaged over Sloan $griz$ filters, is largest immediately after explosion ($\sim0.5$~mag) and falls to $\sim0.145$~mag by day $18$, remaining roughly constant until day $50$. The \sed~uncertainty (grey), measured as root mean square of difference between \sed~\lcs~and \cmf~\lcs~averaged over $griz$ filters, is $\sim0.4$~mag at day $5$, falls to $\sim0.2$~mag by day $20$, then rises again to $\sim0.35$~mag by day $28$ and holds constant until day $50$. The total uncertainty (red), $\sigma_{tot}=\sqrt{\sigma_{\rm emulator}^2+\sigma_{\texttt{SEDONA}}^2}$, ranges from $\sim0.6$~mag just after explosion to a minimum of $\sim0.24$~mag near day $20$, settling at $\sim0.38$~mag at late times.}
    \label{fig:emulator_uncertainty}
\end{figure}

Here we measure the fidelity with which the emulator reproduces SESN \lcs. We use the root mean square (RMS) of the emulated versus true magnitude averaged over $griz$ filters to quantify emulator performance. We find the emulator is least accurate at early epochs, before $\sim18$ days after explosion (see blue function in Figure~\ref{fig:emulator_uncertainty}). These early epochs correspond to the rise phase of the \lc, when the light-curve shape evolves fastest and is most diverse across the grid. Approximately $80\%$ of our grid light curves do not peak until later (median $griz$ peak $\approx25$ days). The RMS of the reconstructed versus \sed~\lc~is $\approx0.5$~mag immediately at 5 days after explosion (when simulations start). The RMS value then decreases to $\approx0.145$~mag by day 18 and remains constant until day 50. Aside from time after explosion, we find that the emulator's performance is not dependent on any other physical parameter. Crucially, the uncertainty of the emulator presented here is comparable to the typical magnitude error in observed SESN \lcs, meaning that it is important to quantify this uncertainty during inference. A significantly larger \lc~grid (possibly exceeding $\sim$10,000 simulated \lcs) would be necessary to train an emulator with sufficiently low uncertainty. 

In addition to measuring the emulator's systematic error in reproducing \sed~\lcs, we also estimate the \sed's systematic error of modeling true SESN \lcs. \sed~assumes the ejecta profile is always in local thermal equilibrium (LTE), dramatically reducing the computational cost of the radiation transfer calculation. However, a realistic RT simulation of SESN \lcs~needs to make relax LTE assumptions to account for relevant phenomena such as the excitation of \he~by photons from \NI~decay \citep{CMFGEN_description}. \lcs~generated from \sed~will be the most realistic at phases where the LTE assumption is most accurate for the SESN ejecta. This is during \lc~peak, and the LTE assumption will be least applicable both immediately after explosion and long after \lc~peak. Immediately after explosion, the ejecta temperature profile will vary until the energy inputted into the ejecta from \NI~decay is equal to the energy emitted by the ejecta at every level. At phases when this equilibrium is being established, the systematic error of \sed~will be large. In contrast, the Arnett model explicitly assumes this holds throughout the evolution of the \lc~(Assumption 5 of \citealt{Arnett1982}; see also \citealt{Khatami2019}). At later phases, the LTE assumption again breaks down as the ejecta will become optically thin. Thus, \sed's systematic error grows again toward late times. To crudely estimate the uncertainty due to LTE assumptions, Y26 compare \sed~simulated \lcs~of nine ejecta models to the same models simulated in \cmf, a code that allows for some non-local thermal properties. We present the RMS of \sed~versus \cmf~\lcs~as a function of time after explosion as the grey function in Figure~\ref{fig:emulator_uncertainty}. The RMS of \cmf~versus \sed~\lcs~is $\sim0.4$mag at $5$ days after explosion and decreases to $\sim0.2$ mag by 20 days post explosion. Then, the RMS again increases to $\sim0.35$mag by $\sim28$ days and holds constant until day $\sim50$. \footnote{Another consequence of making the LTE assumption throughout is that synthetic \lcs~at shorter wavelengths than in the $g$ filter (i.e. $u$ filter) are inaccurate by ~$1$~mag or more. Nonthermal excitations of atoms, which are not modeled in an LTE-only regime, contribute significant flux to the \lc~in wavelengths shorter than optical.} 

We build a new model in the Modular Open Source Fitter for Transients (\mos; \citealt{guillochon2018}) to perform \lc~inference using this emulator. In Sections \ref{sec:fitting_sim} and \ref{sec:fitting_real}, we use \mos~to present the limits and capabilities of performing physical inference on SESN \lcs. Throughout these sections, we report posterior constraints on physical parameters (e.g. \mej~and \gamnn) as their $16$th--$84$th percentile ($1\sigma$) credible intervals, whereas ranges quoted for velocity or mass profiles instead denote the spread across the indicated fraction of ejecta mass. We account for both sources of systematic error in our \mos~fitting pipeline. We fit an analytical function to each of the empirical emulator uncertainty and \sed~uncertainty curves.  The analytical functions for the \sed~uncertainty and emulator uncertainty in terms of the time since explosion ($t$) are presented in Equations~\ref{eq:em_unc} and \ref{eq:sed_unc}, respectively. 

\begin{equation}
    \sigma_{\rm emulator} =  0.125 + 1.70\,\exp(-0.278t)
    \label{eq:em_unc}
\end{equation}

\begin{equation}
\begin{split}
    \sigma_{\texttt{SEDONA}} = 0.36
    +0.048 \exp\left(-\frac{1}{2}\left(\frac{t - 4.33}{0.51}\right)^2\right)\\ 
    -0.15 \exp\left(-\frac{1}{2}\left(\frac{t - 19.01}{3.69}\right)^2\right)\\
    +0.17 \exp\left(-\frac{1}{2}\left(\frac{t - 4.33}{8.05}\right)^2\right)
    \label{eq:sed_unc}
\end{split}
\end{equation}
The total uncertainty $\sigma_{tot} = \sqrt{\sigma^2_{\rm emulator} +  \sigma^2_{\texttt{SEDONA}}}$ is introduced into \mos~as an additional uncertainty term in the likelihood calculation for each draw of the forward model. A version with both uncertainties included is available as a new model \texttt{sesn-sedona} in \mos~\footnote{\url{https://github.com/guillochon/MOSFiT}\\We note that this update, v1.3, represents a major update to \mos. In addition to the ability to fit emulators such as the one presented here, we have also heavily modified user input/output, documentation, and included new features (e.g., fitting in luminosity) as part of this update.}. The resulting total uncertainty (red curve in Figure~\ref{fig:emulator_uncertainty}) ranges from $\sim0.6$~mag at 5 days after explosion to a minimum of $\sim0.24$~mag near day $20$, rising to $\sim0.38$~mag at late times.

\section{Fitting To Simulated Light Curves}
\label{sec:fitting_sim}

In this section, we present results of fitting simulated \sed~\lcs. We highlight how well each of the nine physical parameters can be inferred from full multiband \lc~fitting, and we present notable degeneracies inherent in the fitting process. We then compare the fidelity of physical inference using our emulator to that from using the Arnett model. In every fit presented here, we simultaneously fit to the light curve in every filter. 

\begin{figure}
    \centering
    \includegraphics[width=\linewidth]{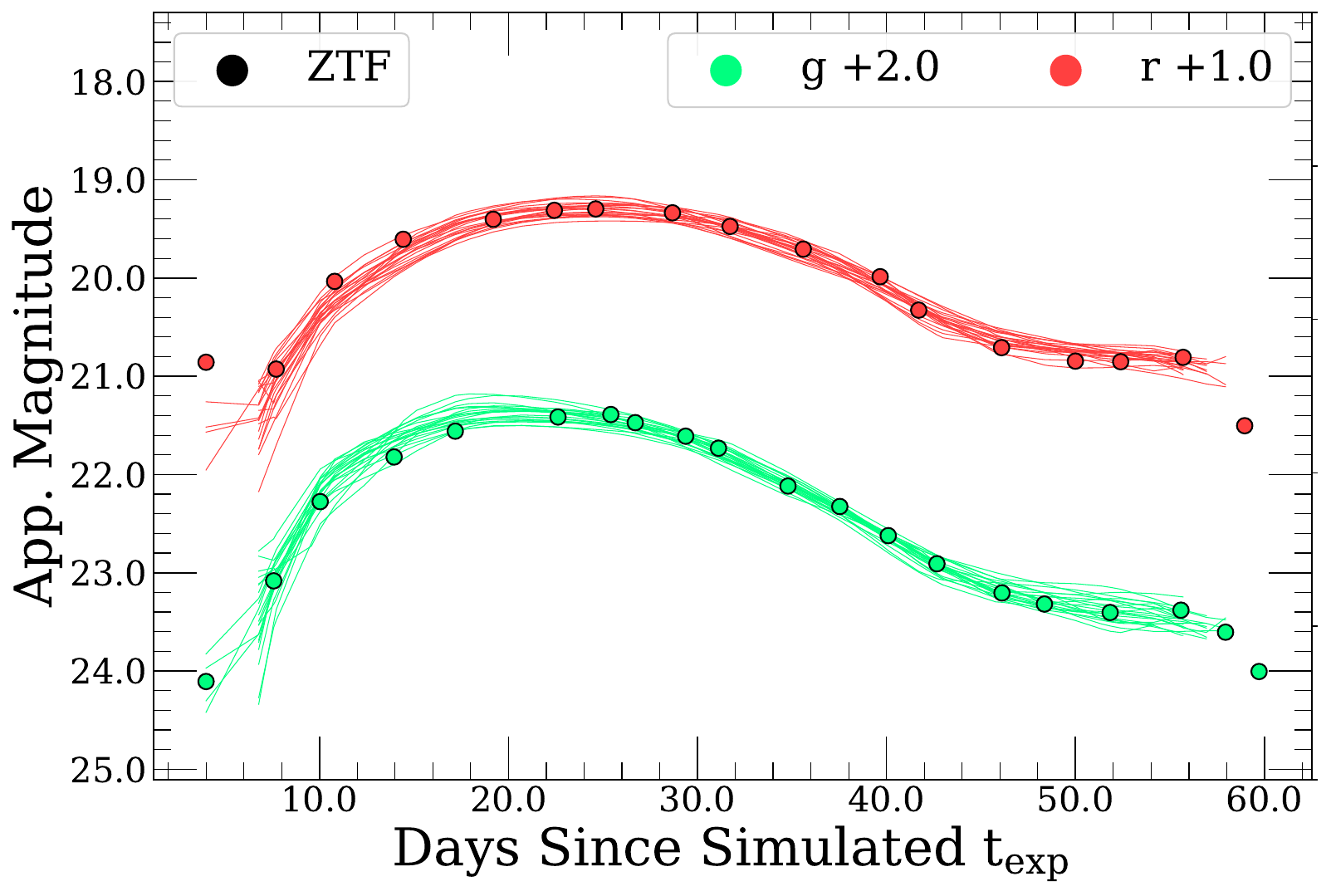}
    \includegraphics[width=\linewidth]{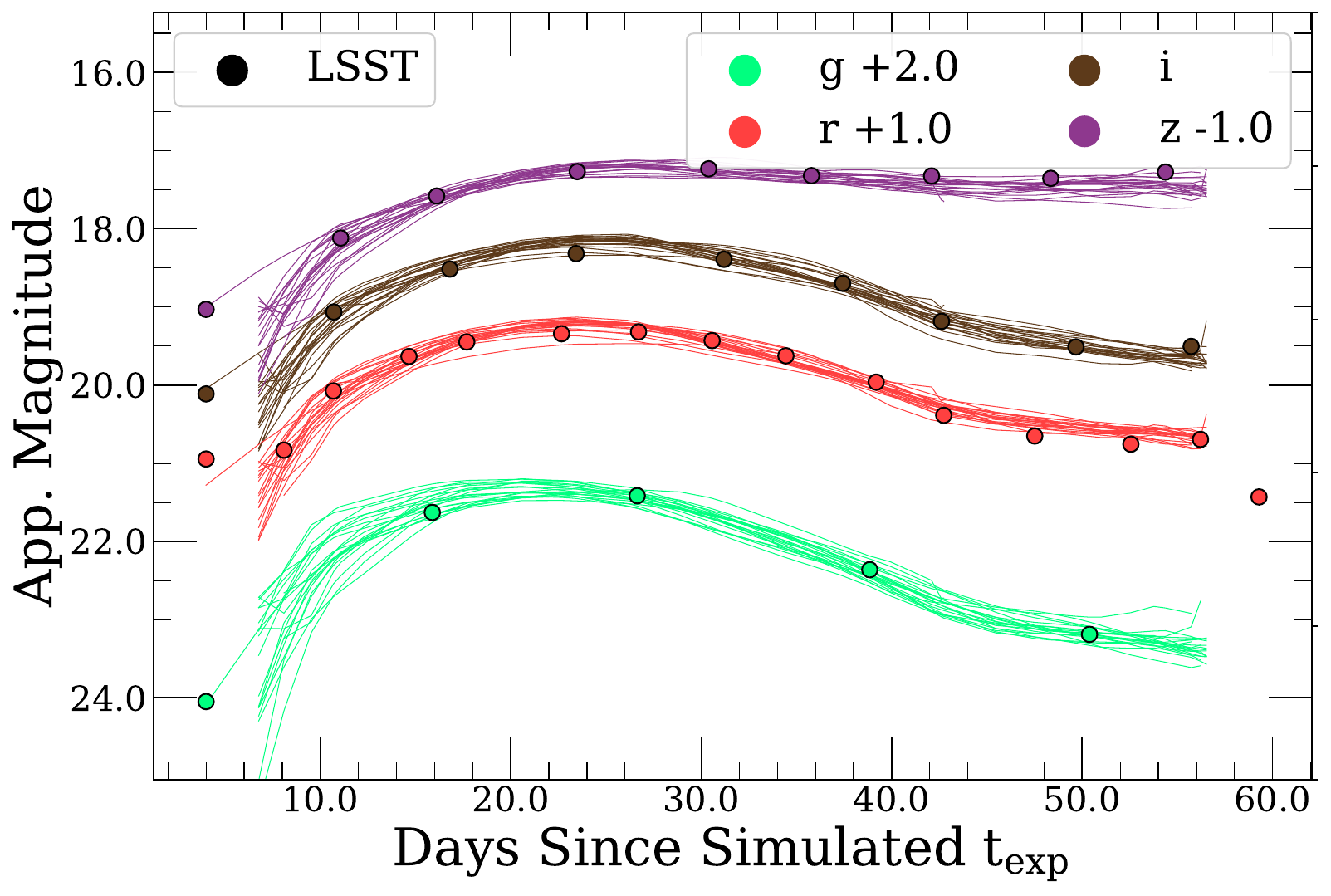}
    \caption{Fits performed to a simulated ZTF-like (top) and LSST-like (bottom) version of the same underlying \lc. }
    \label{fig:sample_0_lc}
\end{figure}

We generate \lcs~with our emulator to match the cadence and filters of observed SESN samples, specifically those from the Bright Transient Survey (BTS; \citealt{Fremling2020, Perley2020}) of the Zwicky Transient Facility (ZTF; \citealt{Bellm2019, Graham2019}) and those from the LSST. ZTF observes the northern sky in the $g$ and $r$ filters with an average cadence of $\sim2$~days in each filter. Beyond ZTF, in the era of the LSST, $\sim$60k new SESNe are expected to be discovered every year \citep{Kessler2019}, marking a dramatic increase in the known SESN population. SESN \lcs~from LSST are expected to be observed in $ugrizy$ filters with typical cadences of 21, 11, 4, 6, 6, and 8 days, respectively \citep{opsim}. To generate \lcs~with realistic magnitude errors, we calculate the expected brightness-dependent signal-to-noise ratio (SNR) of photometric observations for both surveys. For the ZTF \lcs, we calculate the average SNR of the observation as a function of apparent magnitude over all BTS \lcs. For the LSST \lcs, we calculate the expected SNR as a function of apparent magnitude using the method outlined in \cite{lse40}. Then, for both we add noise by sampling from a Gaussian with standard deviation of $\rm 1/{SNR}$ to the \lc~at every phase. Throughout this subsection, we include only the emulator's systematic error (i.e.,  not \sed's systematic error) during the fitting process because the \lcs~we fit to are themselves simulated with \sed. In Figure ~\ref{fig:sample_0_lc}, we show an emulated \lc~with ZTF-like and LSST-like sampling and scatter alongside fits. The simulated \lcs~are placed at 100 Mpc.

Here, we perform inference over a ZTF-like version and an LSST-like version of several emulated SESN \lcs~to map out the fundamental limits of physical inference (see Figure~\ref{fig:sample_0_lc}). We then examine how strongly each parameter can be constrained and which parameters are inherently degenerate in our fit. %We find that fits to the LSST-like versus ZTF-like \lcs~result in similar uncertainties and comparable posterior distributions. 
We find that the fits to the LSST-like vs ZTF-like light curves display similar degeneracies and relative constraints (i.e., He mass is poorly constrained in both surveys). We also find that fitting performance is largely similar across parameter space. The only exception to this is for models with either \mni~$\geq0.2$\msun, the resulting \mni~and \mej~posterior distributions are much broader because the underlying \lc~grid has a paucity of high-\mni~models. We present the LSST-like fit to one typical \lc~and find that the uncertainties and degeneracies in this fit generalize to the majority of fits performed by this emulator over this grid of SESN \lcs.

\subsection{Emulator-Inferred Parameter Constraints}
In Figure~\ref{fig:samp_7_fits_hist}, we present the prior distribution (in orange), the inferred posterior distribution (in blue), and the ground truth (in red) for each physical parameter inferred by our model for the LSST-like light curve. Here we report which parameters are well versus poorly constrained for this representative light curve. The top three rows show the prior and posterior distributions for the nine physical parameters that are directly inputted to the emulator. These have uniform priors, reflecting that the grid is constructed by independently and uniform-randomly sampling from these nine parameters. The bottom row shows $\rm r_{x,95}$, the ``interpretable version'' of the mass distribution parameters: the ratio between the velocity containing $95\%$ of the $\rm x$ mass distribution and the maximum ejecta velocity. As in Y26, because \he~has a negligible influence on the observed \lc, all physical parameters related to the \he~distribution (\mhe, \Rhp, and \gamnh) are poorly constrained. Because the combination of all three of \Dp, $\rm min_{vel}$, and $\rm \Delta_{vel}$ produces the velocity distribution, none of these parameters \textit{independently} describes the velocity distribution, and the constraints on any one of tends to be more loose than the constraint on the overall velocity distribution. Nevertheless, we find that through the combination of constraints on \Dp, $\rm min_{vel}$, and $\rm \Delta_{vel}$, the velocity profile is accurately recovered by this inference method. We discuss velocity and overall ejecta profile fits in detail in the subsequent paragraph. The parameters that describe the \NI~and \opacity~mass distributions (\Rnp~and \Ropp, respectively) are all well recovered by this fit. The median value of the posterior distribution is within $1\sigma$ of the true value for both mass distribution parameters. We also find that strong constraints on the interpretable values \gamnn~and \gamnop~can be placed using our inference method.

In Figure~\ref{fig:samp_7_fits_vel}, we present prior and posterior distributions of ejecta profiles the fits from Figure~\ref{fig:samp_7_fits_hist}. Using draws from the posteriors of the training parameters, we construct distributions of velocity, \he, \NI, and \opacity. Because the \lc~is produced by photons emitted at the photosphere, it most strongly encodes information about the ejecta layers near the photosphere. Layers deep inside the photosphere contribute little to the \lc, because photons emitted there are thermalized and reabsorbed before reaching the photosphere. Layers far outside the photosphere likewise contribute little, since they contain only a small fraction of the radiating material. In the top left panel of Figure~\ref{fig:samp_7_fits_vel}, we show the velocity as a function of the fraction of the total mass held within that velocity. The innermost and outermost velocities are poorly constrained by this fit. However, velocities closer to the center of the ejecta profile, where the photosphere lies, are better constrained.
The top right, bottom left, and bottom right panels of Figure~\ref{fig:samp_7_fits_vel} show the fraction of total \he, \NI, and \opacity~mass as a function of the velocity of the ejecta layer containing that fraction of mass, respectively. In each of these, the posterior mass distribution is both considerably tighter than the prior and overlaps with the true distribution. Unsurprisingly, the \he~distribution posterior is broader than those of \NI~and \opacity. As such, though the precise numbers that parameterize the velocity and mass distributions are not directly strongly constrained (as in Figure~\ref{fig:samp_7_fits_hist}), we emphasize that the velocity and mass distributions are well constrained (as in Figure~\ref{fig:samp_7_fits_vel}).

\subsection{Inferred Parameter Degeneracies}
Next, we explore the parameter degeneracies via the joint posterior distributions. To learn from the joint posterior distributions of the \gamn~parameters, we must first understand the shape of their prior distributions as a potential bias. The \gamn~parameters do not trivially have uniform priors because these parameters are computed for each draw, rather than uniformly sampled. When \gamnn~is large, the velocity containing most of the \NI~is large (i.e., the \NI~is mixed out). This can happen either because the \NI~profile is inherently broad or because the velocity distribution is shallow, such that the outermost velocities are not much larger than median velocities (i.e., \Dp~is close to $0$). Similarly, a smaller \gamnn~can be produced by either a less inherently mixed \NI~profile, or by a steeper velocity distribution (i.e. \Dp~close to $1$). However, the influence of the range of \NI~profiles on the \lc~is not as strong as that of the range of velocity profiles in our grid. Therefore, the lowest values of \gamnn~are largely only produced by velocity profiles with \Dp~close to one, whereas varying \Rnp~when \Dp~is close to one largely does not impact \gamnn. This effect is also true for the \he~and \opacity~profiles. As such, a SESN ejecta profile with a small \gamnn~would also necessarily have a small \gamnh~and \gamnop; small \gamnh, \gamnn, and \gamnop~values are typically only produced with a steep velocity profile. Larger \gamnn~and \gamnop~can correspond to a wider range of \Dp~values. 

\begin{figure*}
    \centering
    \includegraphics[width=\linewidth]{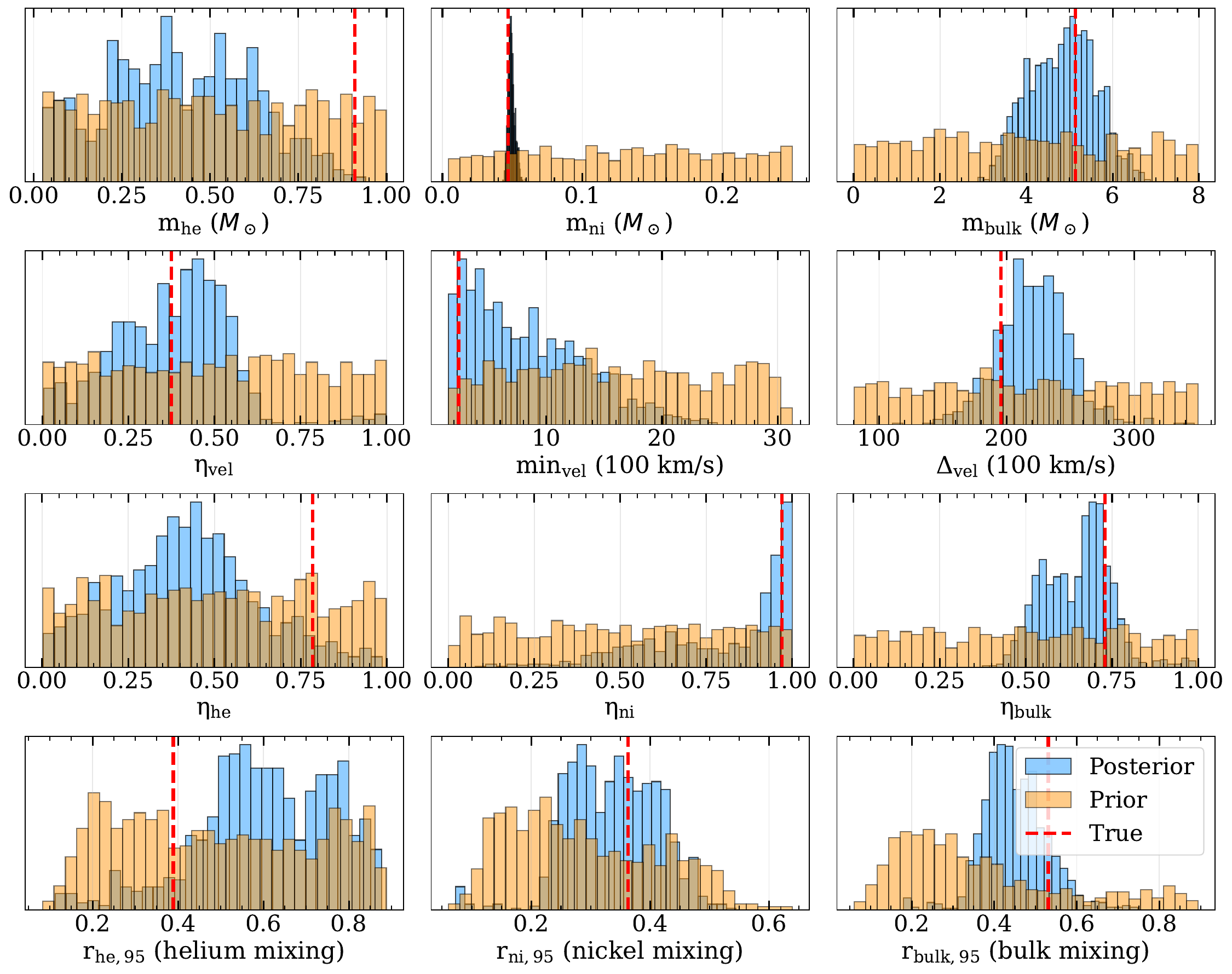}
    \caption{Marginal posteriors of inferred physical parameters for a model in our grid (blue) alongside prior distributions (orange). The ground-truth values are shown in red. Interpretable ejecta distribution parameters (i.e. \gamnn~and \gamnop) are better inferred than latent variables. All physical parameters except for those related to \he~are reproduced by the fit.}
    \label{fig:samp_7_fits_hist}
\end{figure*}

\begin{figure*}
    \centering
    \includegraphics[width=\linewidth]{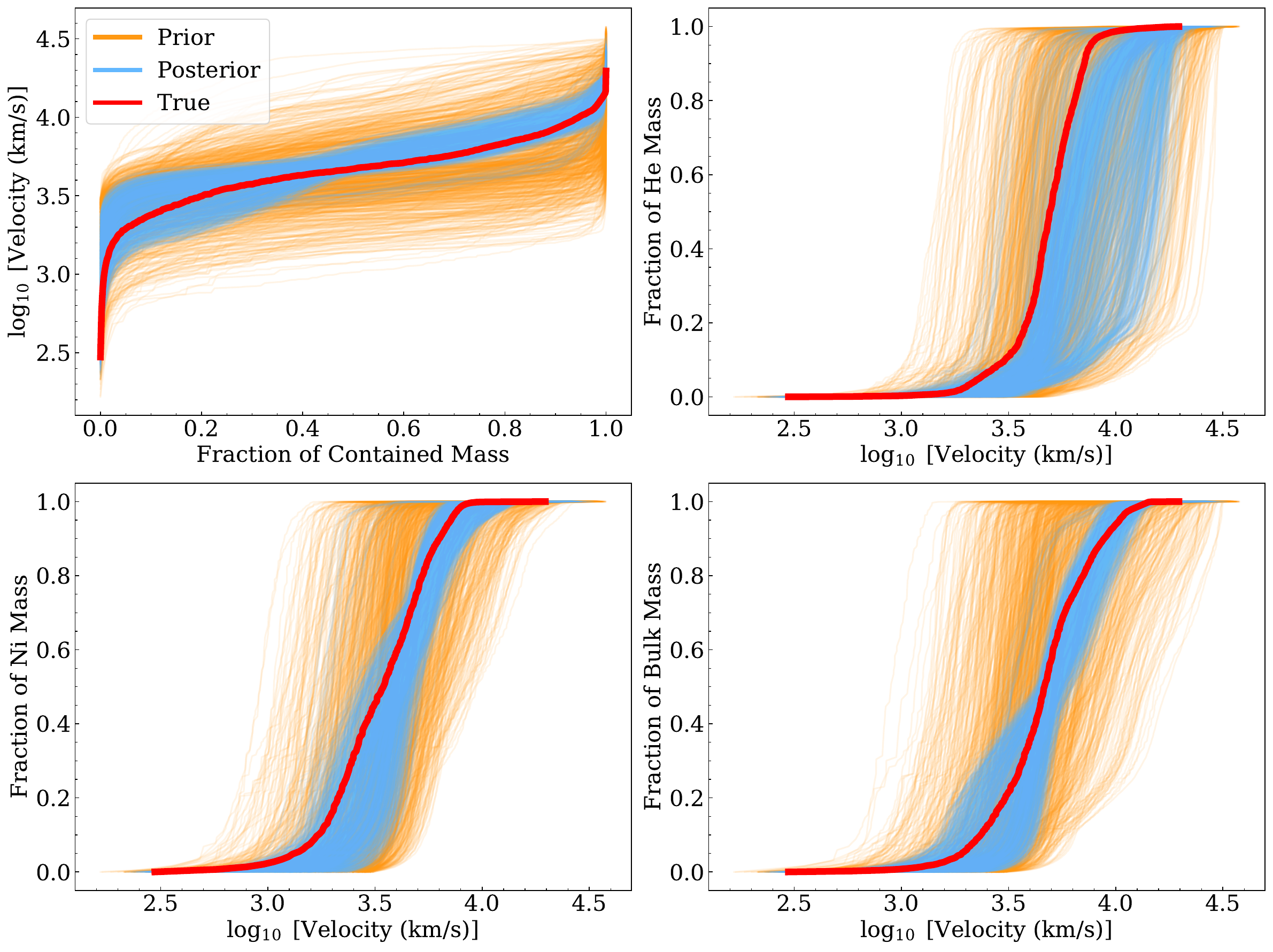}
    \caption{Inferred ejecta profile distributions for a model in our grid.  Prior distributions are shown in orange, posterior distributions are shown in blue, and ground-truth values are shown in red. Most of the velocity profile (top left panel) is recovered, except for velocities of the innermost and outermost layers. The same trend is propagated to to the \he~(top right), \NI~(bottom left), and \opacity~(bottom right) mass distributions.}
    \label{fig:samp_7_fits_vel}
\end{figure*}

\begin{figure*}
    \centering
    \includegraphics[width=\linewidth]{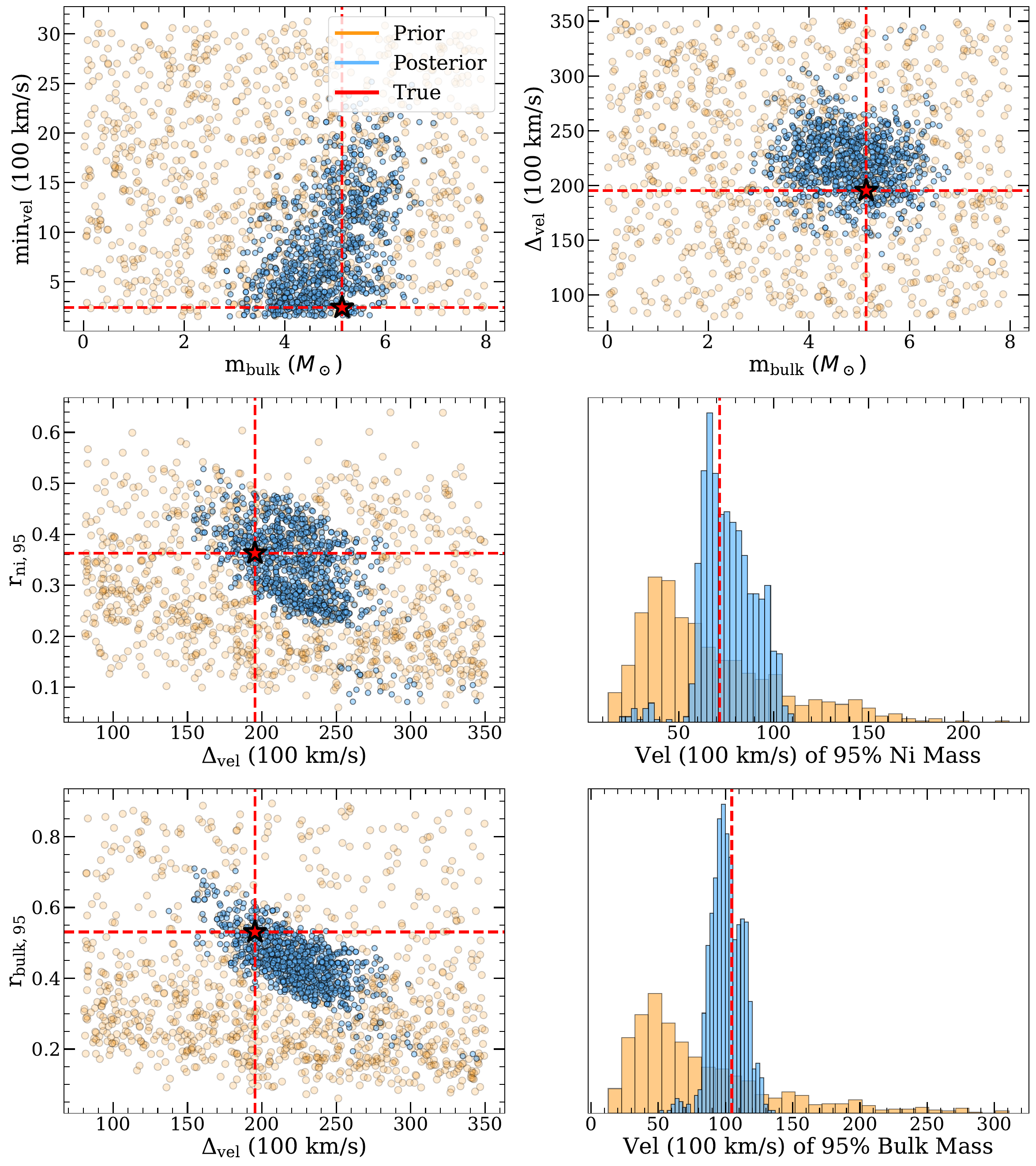}
\caption{Joint distributions of \mop~with $\rm min_{vel}$ (top left), \mop~with $\rm \Delta_{vel}$ (top right), \gamnn~with $\rm \Delta_{vel}$ (middle left), and \gamnop~with $\rm \Delta_{vel}$ (bottom left); the velocity containing $95\%$ of \NI~mass (middle right) and  the velocity containing $95\%$ of \opacity~mass (bottom right) are shown here. Prior distributions are shown in orange, posterior distributions are shown in blue, and ground-truth values are shown in red. }
    \label{fig:samp_7_fits_pairs}
\end{figure*}

In Figure~\ref{fig:samp_7_fits_pairs}, we present the joint posterior distributions of mass and velocity profile of the \lc~fit from Figures~\ref{fig:samp_7_fits_hist} and \ref{fig:samp_7_fits_vel}. The joint distribution of \mop~with $\rm min_{vel}$ is weakly correlated, where larger \mop~correlated with larger $\rm min_{vel}$ (top left panel of Figure~\ref{fig:samp_7_fits_pairs}). This correlation is reminiscent of the Arnett-like degeneracy between \mej~and \vej, where those two parameters are positively correlated. However, we find \mop~and $\rm \Delta_{vel}$ are  uncorrelated (top right panel of Figure~\ref{fig:samp_7_fits_pairs}). In our fitting method, the degeneracy between ejecta mass and characteristic ejecta velocity that is implicit in the Arnett model is far weaker. We find instead that the values $\rm \Delta_{vel}$ and \gamnn~are strongly anticorrelated (see middle row of Figure~\ref{fig:samp_7_fits_pairs}). This is because for samples where $\rm \Delta_{vel}$ is increased, a nearly identical \lc~can be produced by forcing \NI~to be less mixed into the outer layers. We find a similar anticorrelation between $\rm \Delta_{vel}$ and \gamnop~(see bottom row of Figure~\ref{fig:samp_7_fits_pairs}). We find that the fit is able to constrain the velocity containing $95\%$ of \NI~mass in the ejecta and the velocity containing $95\%$ of the \opacity~mass (see right panels of bottom two rows in Figure~\ref{fig:samp_7_fits_pairs}).

We note here that our discussion of degeneracies may be specific to our emulator and our grid of \lcs. Our emulator implicitly assumes that the relative abundances of the elements \textit{within} the \opacity~(carbon, oxygen, neon, etc.) match those in the \cite{Woosley2021} grid. Y26 show that $\sim50\%$ changes to these abundances do not appreciably alter the \lc; however, we cannot rule out that dramatically different compositions would meaningfully impact it. 

\subsection{Comparing to Arnett}
\begin{figure*}
    \includegraphics[width=\linewidth]{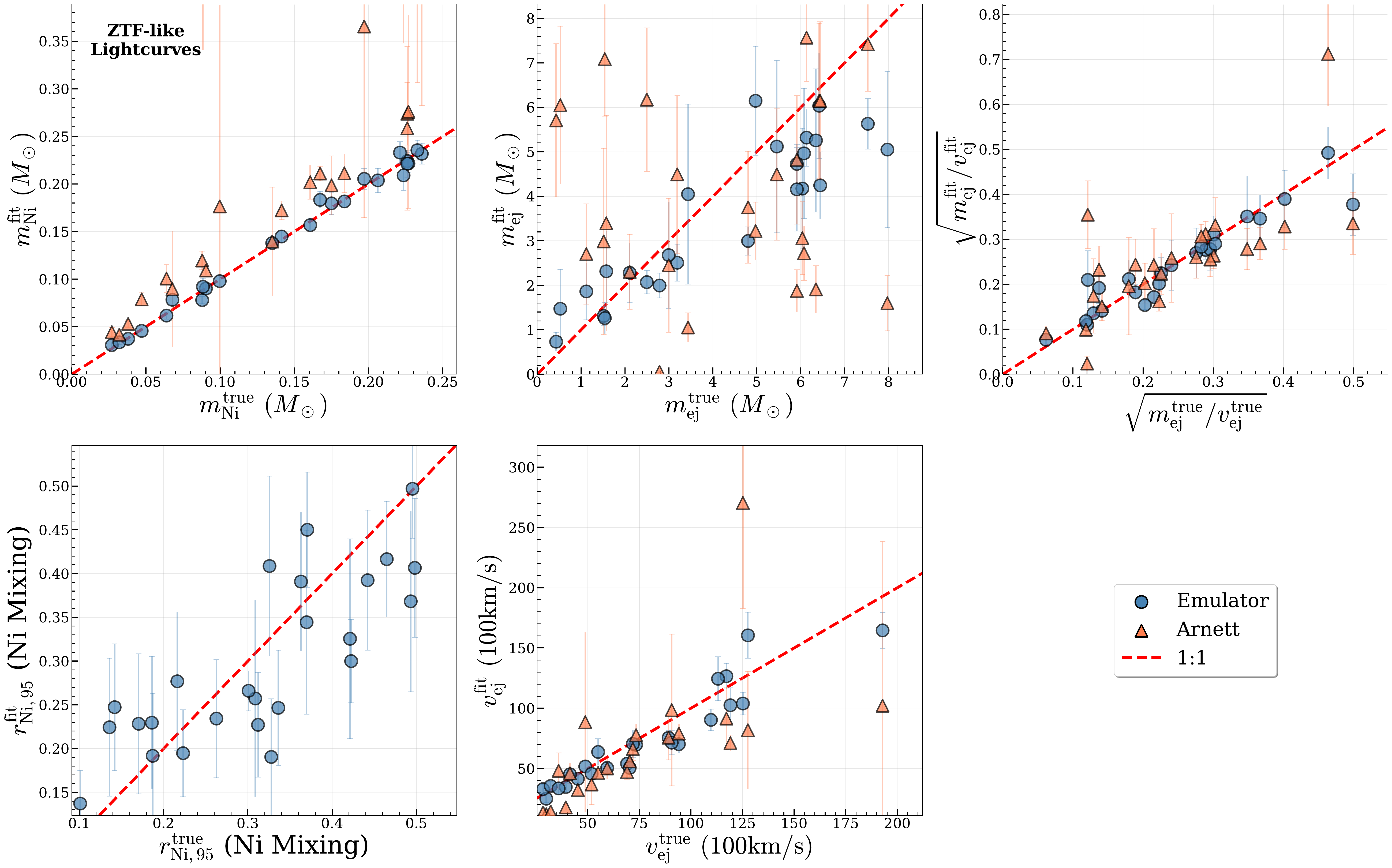}
    \caption{Recovery of physical parameters for simulated ZTF-like \lcs: \mni~(top left), \mej~(top center), and the degenerate combination $\rm \sqrt{m_{ej}/v_{ej}}$ (top right); \gamnn~(\NI~mixing; bottom left) and \vej~(bottom center). Emulator fits are shown as blue circles and Arnett fits as orange triangles, with the dashed $1\!:\!1$ line for reference. The Arnett fits adopt the calibrated opacity $\kappa=0.054~\rm cm^2~g^{-1}$ (see text). Some Arnett-predicted \vej~values fall outside the vertical axis range.}
    \label{fig:ZTF_parameter_recovery}
\end{figure*}

\begin{figure*}
    \centering
    \includegraphics[width=\linewidth]{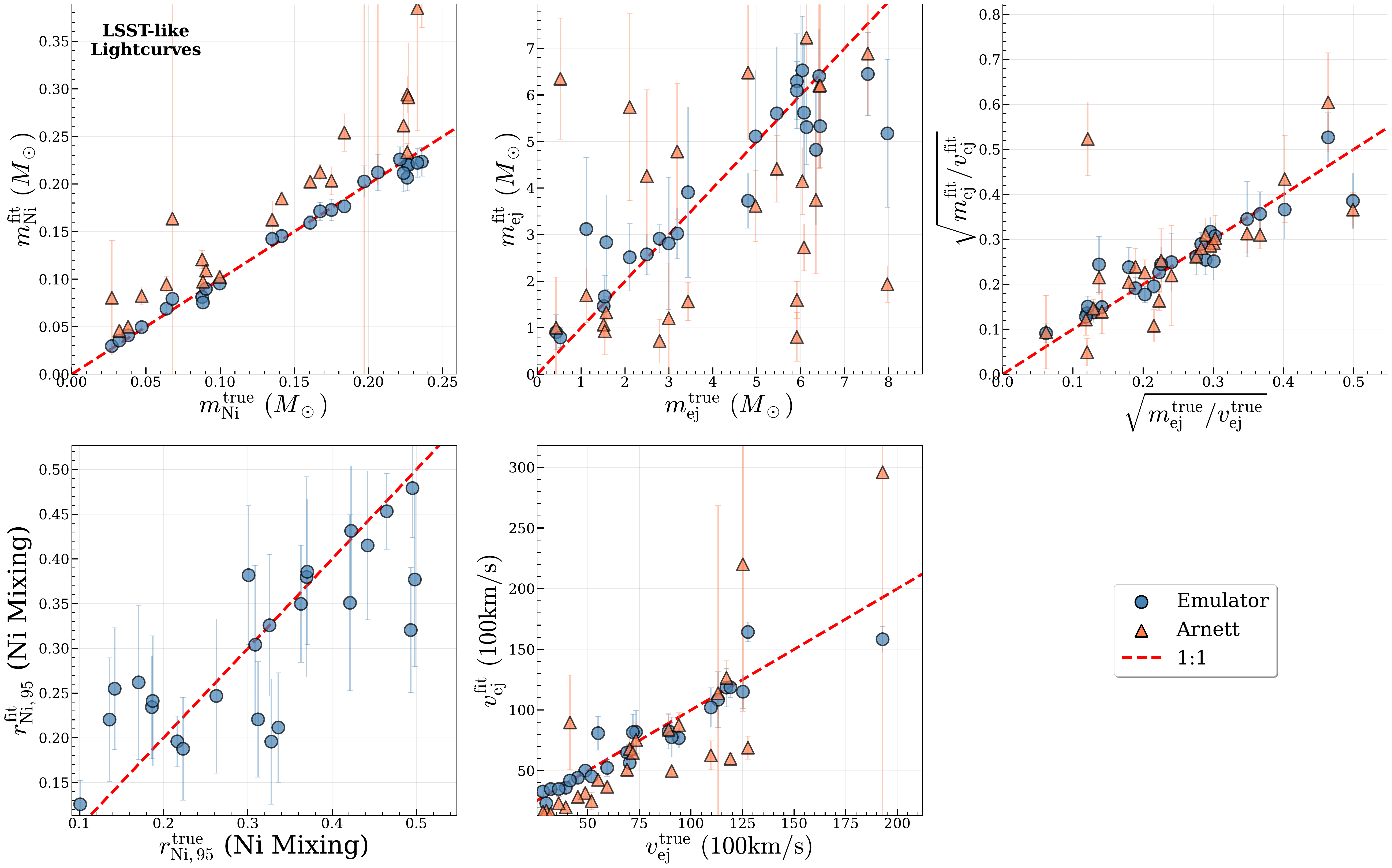}
    \caption{Recovery of physical parameters for simulated LSST-like \lcs: \mni~(top left), \mej~(top center), and the degenerate combination $\rm \sqrt{m_{ej}/v_{ej}}$ (top right); \gamnn~(\NI~mixing; bottom left) and \vej~(bottom center). Emulator fits are shown as blue circles and Arnett fits as orange triangles, with the dashed $1\!:\!1$ line for reference. The Arnett fits adopt the calibrated opacity $\kappa=0.054~\rm cm^2~g^{-1}$ (see text).}
    \label{fig:LSST_parameter_recovery}
\end{figure*}

Here, we characterize how well physical parameters can be constrained by our emulator over a range of ejecta profile parameters. In Figures~\ref{fig:ZTF_parameter_recovery} and \ref{fig:LSST_parameter_recovery}, we show how well \mni, \mej, \vej, $\rm\sqrt{m_{ej}/v_{ej}}$, and \gamnn~are recovered from inference on LSST-like and ZTF-like \lcs. In the Arnett model, the ejecta profile is assumed to be a homologously expanding sphere of uniform density. The kinetic energy of such a sphere is 
\begin{equation}
    KE = \frac{3}{10}m_{ej}v_{ej}^2,
\end{equation}
where \mej~is the ejecta mass, and \vej~is the Arnett characteristic ejecta velocity. Because \vej~is a characteristic velocity measure of the ejecta and does not obviously correspond to any single velocity statistic (i.e., a weighted mean) in our models. To compare emulator-fitted velocity to the Arnett-fitted velocity, we calculate which velocity percentile in our grid of models corresponds most closely to the Arnett characteristic ejecta velocity. We find that the $66$th percentile velocity of an ejecta profile most closely corresponds to the Arnett characteristic ejecta velocity across our simulation grid. We therefore take $\rm \sqrt\frac{10E}{3M}$ as the ``true'' ejecta velocity and the $66$th percentile velocity as the emulator-fitted ejecta velocity to compare to the Arnett-fitted characteristic ejecta velocity.

We fit to each simulated \lc~with the Arnett model using \mos~with the default constant gray opacity of $\kappa=0.2~\rm cm^2~g^{-1}$. With this value, the Arnett-inferred $\rm\sqrt{m_{ej}/v_{ej}}$ is systematically offset below its true value by a factor of $\sim 2$. Because fitting at an incorrect opacity rescales the inferred $\rm \sqrt{m_{ej}/v_{ej}}$ by $\sqrt{\kappa_{\rm true}/\kappa_{\rm assumed}}$, this offset implies an effective opacity roughly a factor of four lower. We therefore re-fit the Arnett model with $\kappa=0.054~\rm cm^2~g^{-1}$ (a value obtained by fitting the average offset). This new, effective opacity removes the offset, and we proceed with using this new value in all experiments below. We note that this value is consistent with the range typically explored in the literature (e.g., \citealt{wheeler2015analysis,taddia2018carnegie}).

We find that our emulator recovers \mni, \mej, and \vej~individually more accurately than the Arnett model for both surveys. We quantify goodness of fit using the root mean squared (RMS) of the difference between fitted and true values. For \mni, the emulator achieves an RMS of $0.008$~\msun~(LSST) and $0.007$~\msun~(ZTF), versus $0.33$ and $0.70$~\msun~for the Arnett model. The emulator error is uniform across the prior range of \mni, whereas the Arnett model systematically overpredicts \mni\, by $\sim80\%$ on average for LSST-like \lcs~and by nearly a factor of three for the ZTF-like \lcs. The emulator recovers \mej~with an RMS of $0.9$~\msun~(LSST) and $1.2$~\msun~(ZTF), versus $2.6$ and $3.0$~\msun~for the Arnett model. The Arnett inference is strikingly poor for \mej, returning essentially draws from the prior. For \vej, the emulator reaches an RMS of $\sim1340~\rm km~s^{-1}$ (LSST) and $\sim1450~\rm km~s^{-1}$ (ZTF), versus $\sim3800$ and $\sim17000~\rm km~s^{-1}$ for the Arnett model. The exceptionally large ZTF Arnett \vej~RMS is driven by a few catastrophic fits in which the mass--velocity degeneracy drives \vej~toward the high-velocity edge of the prior ($\sim7\times10^4~\rm km~s^{-1}$). Excluding these outliers lowers the RMS to $\sim2900~\rm km~s^{-1}$. On the degenerate combination $\rm \sqrt{m_{ej}/v_{ej}}$ (top-right panels of Figures~\ref{fig:ZTF_parameter_recovery} and \ref{fig:LSST_parameter_recovery}), our emulator still outperforms Arnett. The emulator recovers it with an RMS of $0.04$ and the Arnett model with an RMS of $0.10$ (LSST) and $0.09$ (ZTF). While by definition the Arnett model can only constrain this degenerate mass-velocity combination, our emulator additionally separates \mej~and the velocity profile of the ejecta. Our emulator can also recover the \NI~mixing for both surveys (bottom-left panels of Figures~\ref{fig:ZTF_parameter_recovery} and \ref{fig:LSST_parameter_recovery}). It recovers \gamnn~with an RMS of $0.07$ for both LSST-like and ZTF-like \lcs, with uniform scatter across the range of \NI~mixing explored here.

Fits performed to ZTF-like and LSST-like \lcs~result in similar accuracy in the fitted physical parameters, with ZTF-like \lcs~recovering \mni~marginally better (RMS $0.007$ versus $0.008$~\msun) and LSST-like \lcs~recovering \mej~better (RMS $0.9$ versus $1.2$~\msun). The color evolution of SESN \lcs~encodes physical information that is well captured by the broad LSST filter set, which may aid the recovery of \mej~in particular. This result is encouraging for the prospect of physical inference on the upcoming sample of LSST SESN \lcs.

\section{Fitting to Real Light Curves}
\label{sec:fitting_real}

\begin{figure*}
    \includegraphics[width=0.49\linewidth]{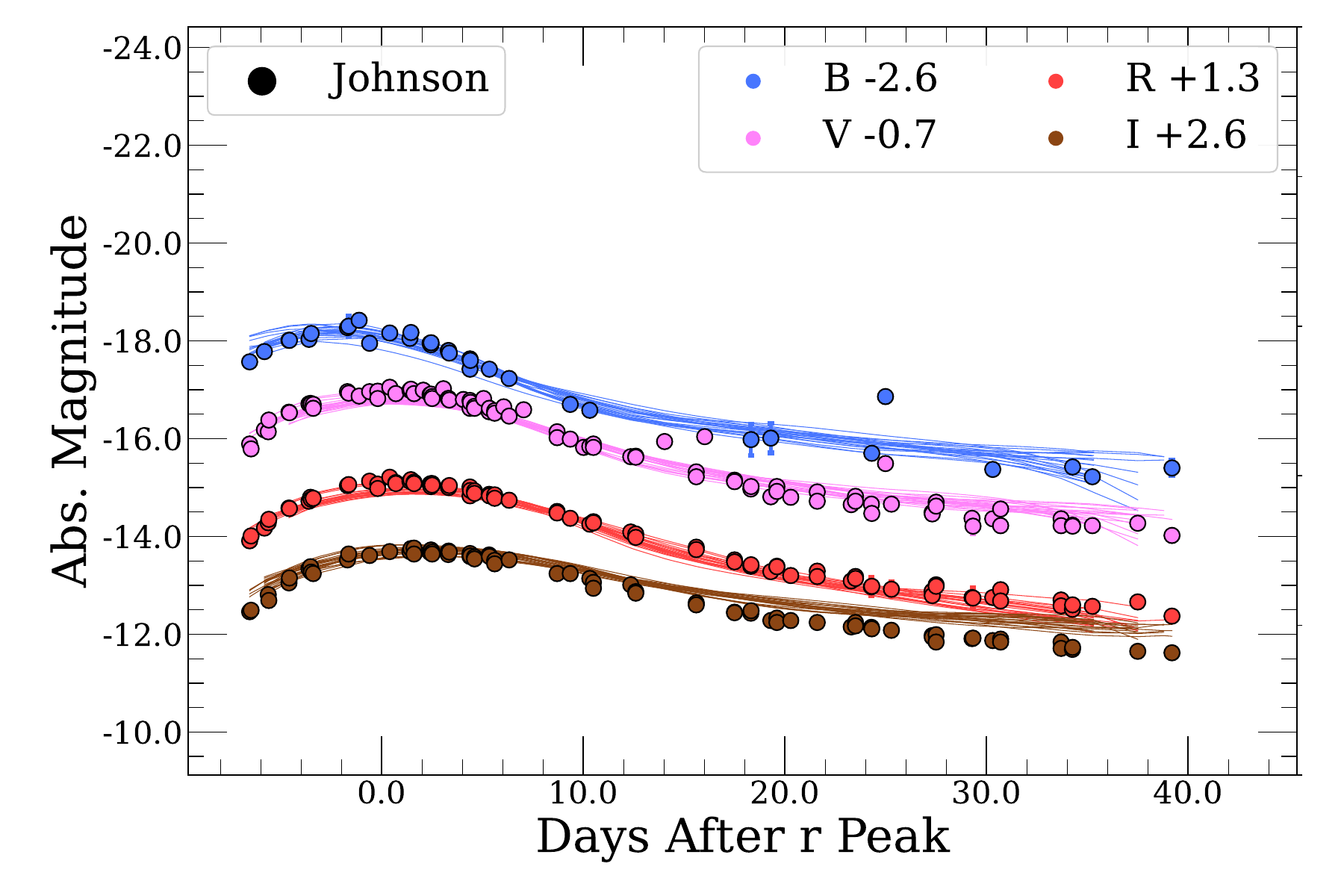}
    \includegraphics[width=0.49\linewidth]{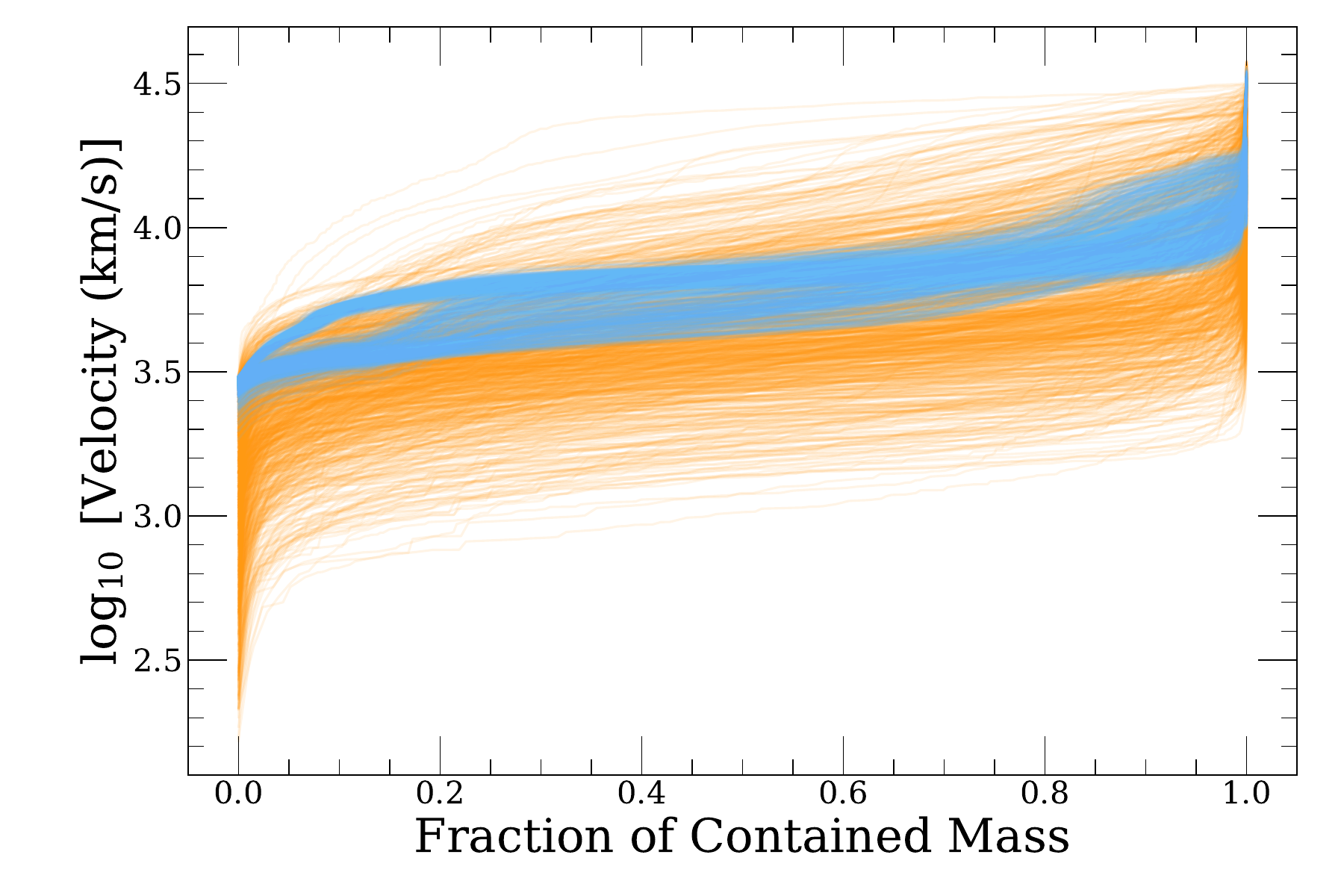}
    \includegraphics[width=\linewidth]{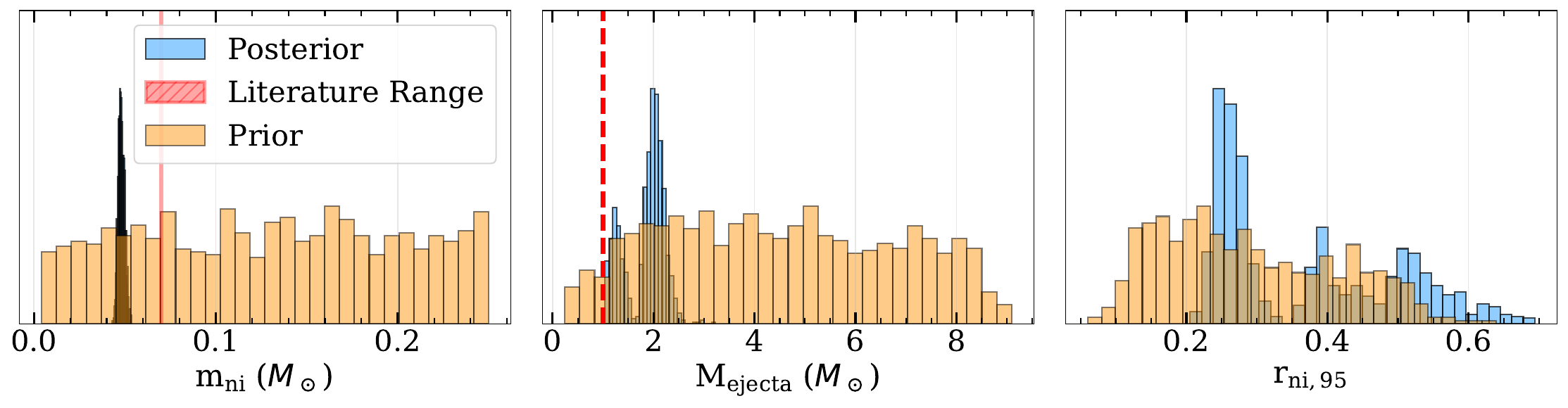}
    
    \caption{Inferred physical parameters, \lc~overlays, and inferred ejecta profiles for SN~1994I from our custom model on \mos~with our emulator. Literature ranges from \cite{Iwamoto1994, Nomoto1994, Young1995}. }
    \label{fig:94I_lc_fits}
\end{figure*}

\begin{figure*}
    \includegraphics[width=0.49\linewidth]{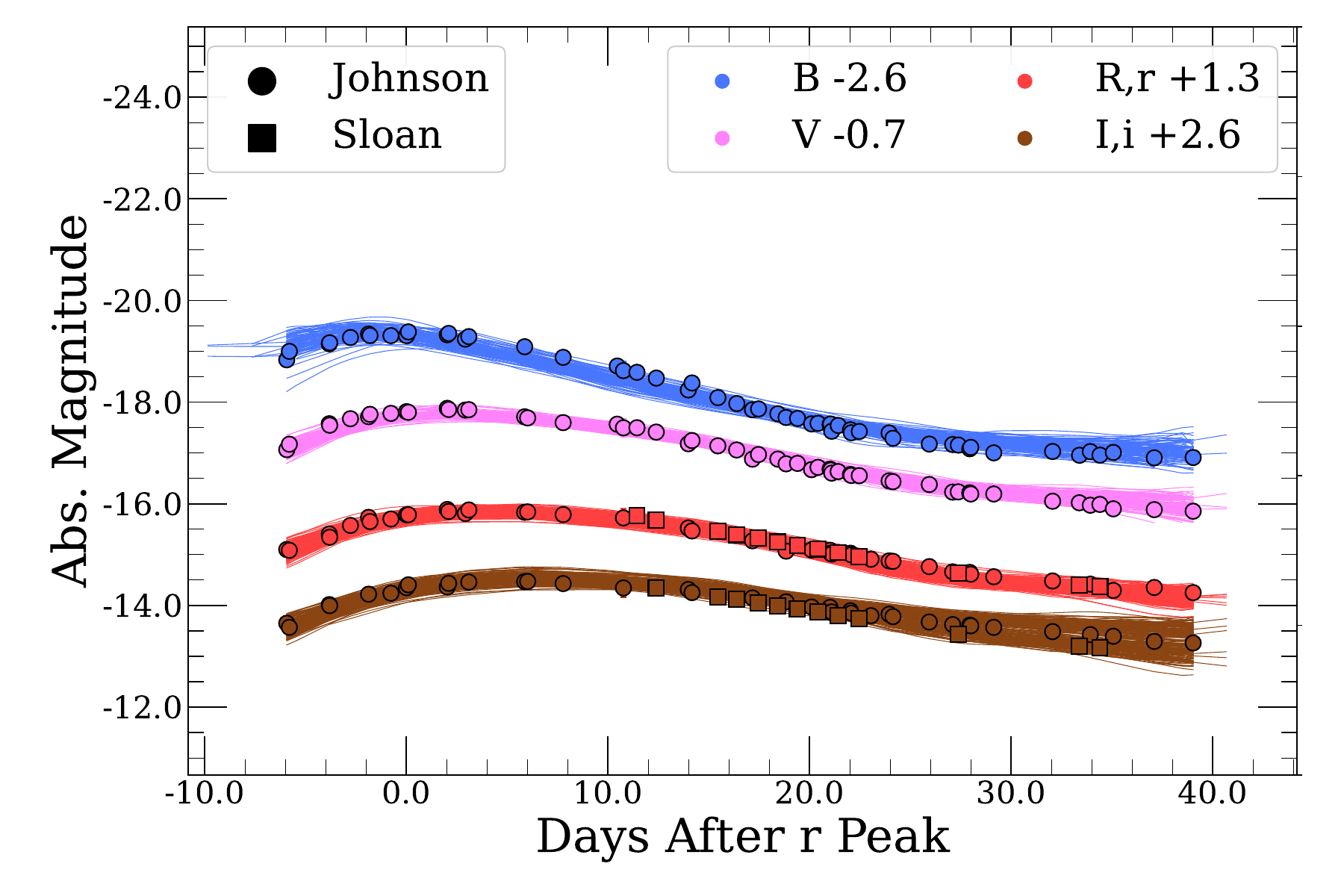}
    \includegraphics[width=0.49\linewidth]{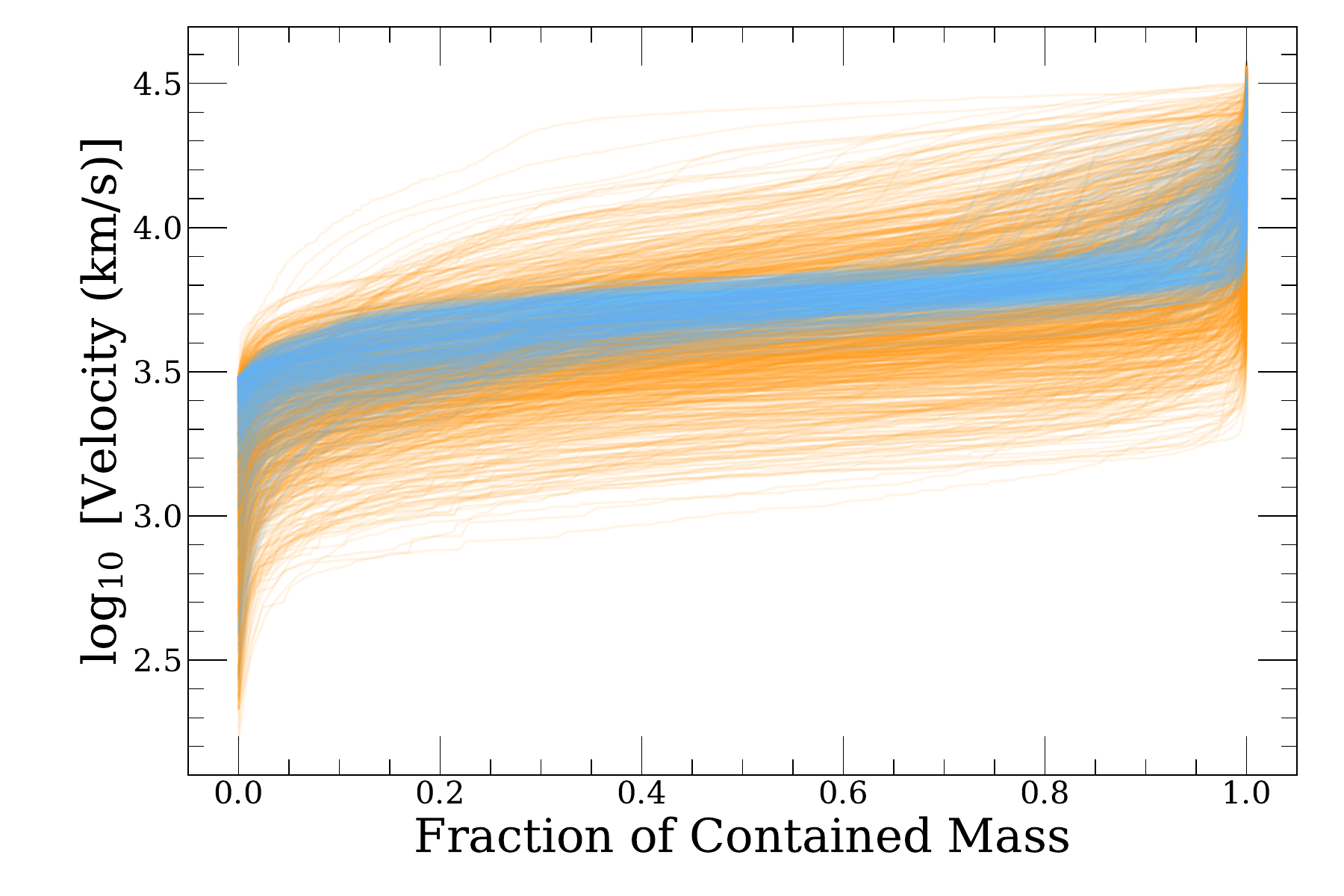}
    \includegraphics[width=\linewidth]{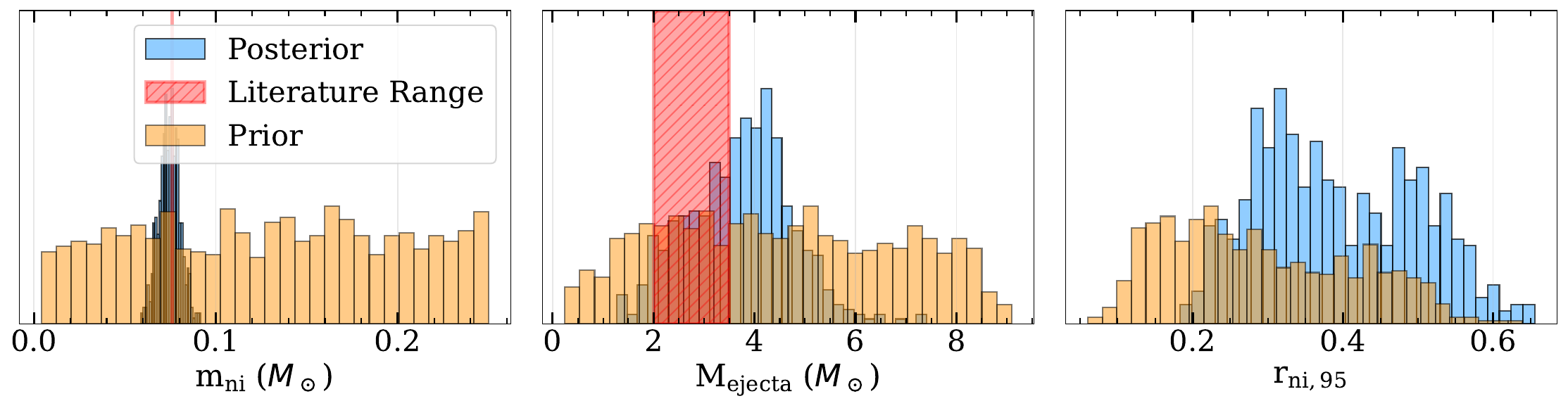}
    
    \caption{Inferred physical parameters, \lc~overlays, and inferred ejecta profiles for SN~2007gr from our custom model on \mos~with our emulator. Literature ranges from \cite{Valenti2008, Hunter2009, Mazzali2010}}
    \label{fig:07gr_lc_fits}
\end{figure*}

\begin{figure*}
    \includegraphics[width=0.49\linewidth]{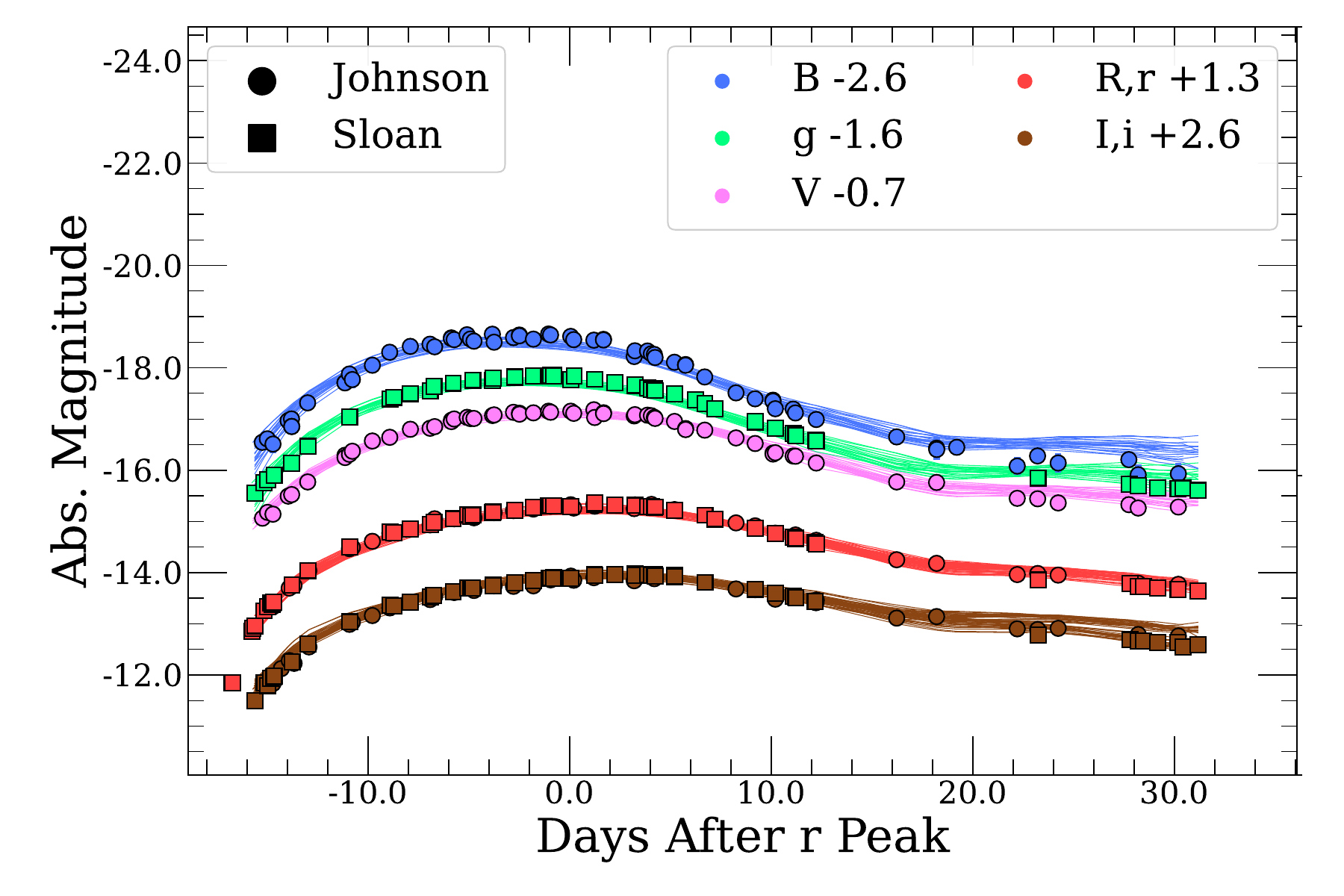}
    \includegraphics[width=0.49\linewidth]{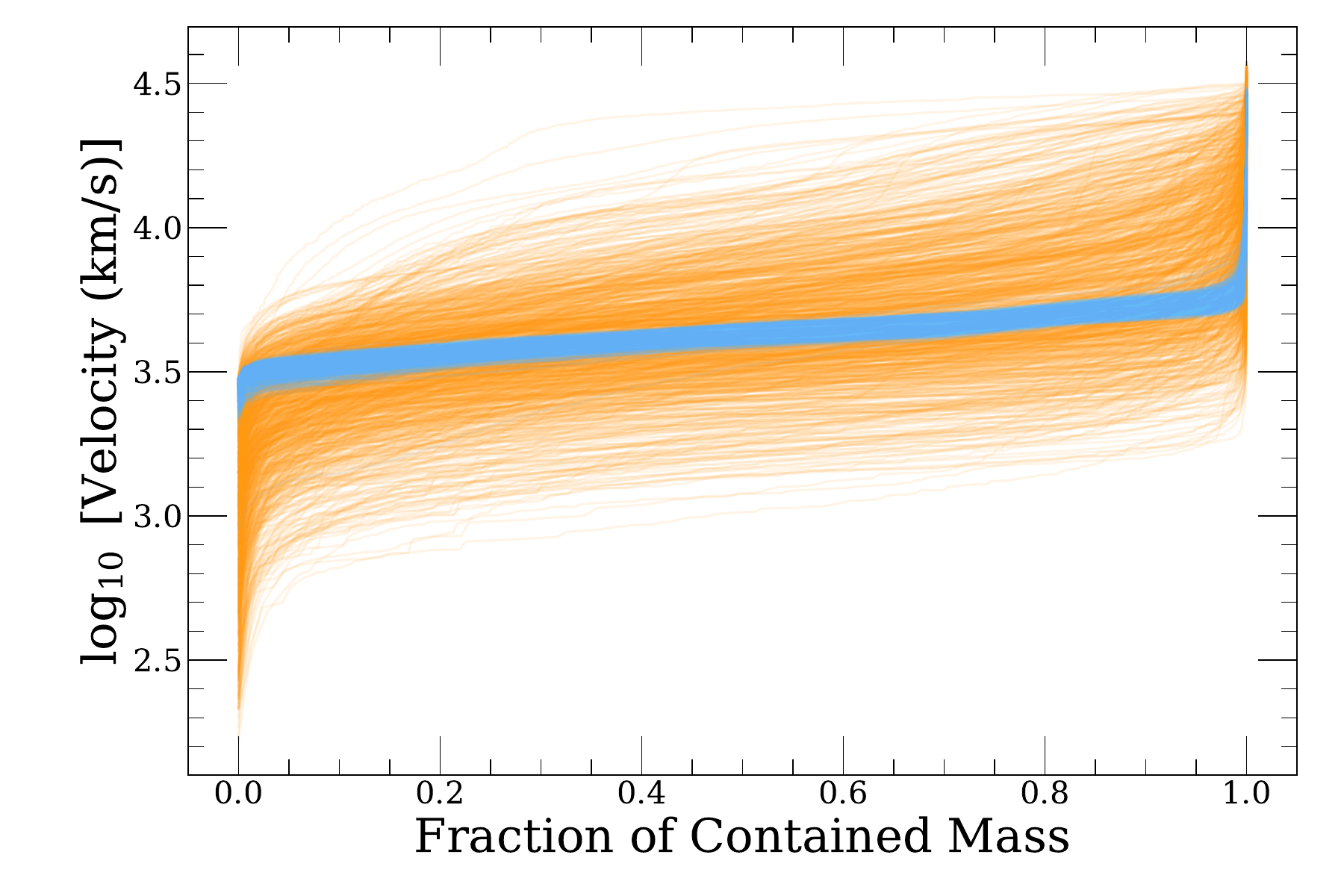}
    \includegraphics[width=\linewidth]{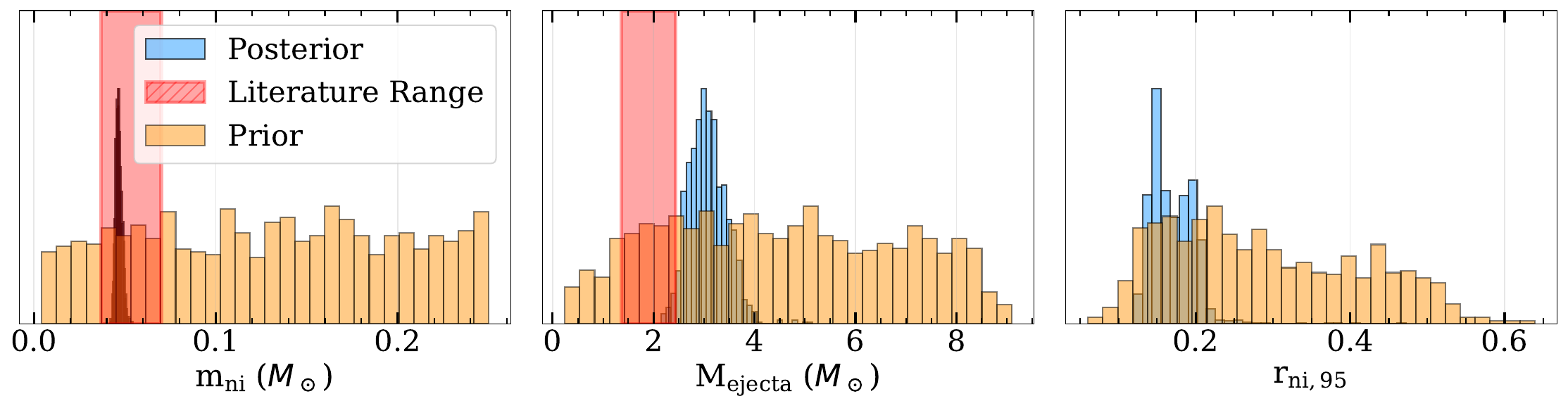}
    
    \caption{Inferred physical parameters, \lc~overlays, and inferred ejecta profiles for iPTF13bvn from our custom model on \mos~with our emulator. Literature ranges from \cite{Fremling2014}}
    \label{fig:13bvn_lc_fits}
\end{figure*}

In this section, we present fits to real \lcs~using our \sed~emulator. Y26 presented nearest-neighbor matches from within a grid of \lcs~generated by \sed to the multiband \lcs~of three well-studied SESNe: SN~1994I, SN~2007gr, and iPTF13bvn. Even though they only perform template matching (rather than true Bayesian inference), they find simulated \lcs~that broadly reproduce the magnitude, timescale, and color of all three observed \lcs. With our trained emulator, we fit to the same three \lcs, where we obtain their photometry from \cite{Khakpash2024}. In Figures~\ref{fig:94I_lc_fits}, \ref{fig:07gr_lc_fits}, and \ref{fig:13bvn_lc_fits}, we show the  best-fit \lcs~and their corresponding best-fit physical parameters as found by \mos~for SN~1994I, SN~2007gr, iPTF13bvn, respectively. We fit to SN~1994I and iPTF13bvn \lcs~using the \texttt{dynesty} sampler \citep{speagle2020dynesty} and to the SN~2007gr \lc~using the \texttt{emcee} sampler \citep{foreman2013emcee}. In all three cases, our \lc~fits here match the observed \lcs. Consequently we find broadly the same values for \mni, \mej, and the velocity profile here, as was found by Y26. Beyond \mni, \mej, and the velocity profile, our inference also constrains the degree of \NI~mixing (\gamnn). We compare all values to inferences from the literature. Throughout this work \mej~is defined as the sum of \mni, \mop, and \mhe. 

%talk about 1994I
We first discuss the \lc~of SN~1994I, a type Ic SN detected in the nearby galaxy M51, at luminosity distance of $\sim7$Mpc \citep{Schmidt1994}. The SN has been modeled extensively, with most studies suggesting that is was powered by $\sim0.07$\msun~of \mni~and that the total ejecta has mass $\sim1$\msun~\citep{Iwamoto1994, Nomoto1994, Young1995}. These estimates were derived by matching the peak luminosity of the \lc. Mixing of radioactive material has also been modeled in the literature. \cite{Sauer2006} modeled the early photospheric spectra and found a substantial abundance of iron-group elements in the outer, high-velocity layers. The spectral features suggested that \NI~is mixed outwards into the oxygen-rich ejecta. Similarly, \cite{Yoon2019} concluded that SN~1994I is heavily mixed based on pre-peak colors of the light curve. Following the work of \cite{Sauer2006}, we use a total line of sight extinction of $E(B-V)=0.3$~mag in our fitting procedure. 

With our emulator, we infer \mni~of $0.048^{+0.002}_{-0.002}$\msun, \mej~of $1.954^{+0.228}_{-0.669}$\msun, and \gamnn~of $0.352^{+0.180}_{-0.100}$. While the median of the \gamnn~posterior ($0.352$) lies in the upper third of our grid prior, the posterior is skewed toward high values, with a tail extending to \gamnn~$\sim0.7$ — well beyond the bulk of the prior. This indicates \NI~mixed substantially into the outer ejecta layers, consistent with the heavy outward mixing inferred by \cite{Sauer2006} and \cite{Yoon2019}. We also find that the bulk of the ejecta (the middle $\sim$90\% of the mass) has velocity $\sim10^{3.6-4.0}$, with only the innermost few percent falling below $10^{3.5}~\rm km~s^{-1}$ (Figure~\ref{fig:94I_lc_fits}). Interestingly, our estimate of \mni~is lower than the $\sim0.07$\msun~that is consistently found by previous \lc~studies. Recently, however, \cite{afsariardchi2021} measured the \mni~using the tail of the light curve, finding a mass of $\simeq0.048\,M_\odot$--identical to our inferred value. This is only one of two objects in their sample for which the inferred dimensionless $\beta$ parameter (an approximate ``correction'' for \NI~powered \lcs~with arbitrarily mixed heating distributions) is close to zero, and they suggest that an additional power source may be the root cause. Our model suggests that a high degree of mixing may account for the discrepancy between the peak luminosity and tail, instead of an additional power source. We note that the \mej~posterior for SN~1994I is bimodal, with a dominant family of solutions near \mej$\approx2$\msun~and a secondary family near \mej$\approx1.2$\msun. The two families share a nearly identical \mni~but differ in their velocity profiles, with a narrower $\rm \Delta_{vel}$ for the lower-mass solution, and both families reproduce the observed \lc~comparably well ($\Delta\ln\mathcal{L}\approx0.8$), a manifestation of the residual ejecta mass--velocity degeneracy discussed in Section~\ref{sec:fitting_sim}. Furthermore, within the \mej~posterior distribution, \mop~is constrained to $1.00^{+0.23}_{-0.33}$\msun~with a single peak. The bimodality in the \mej~posterior comes from bimodality in the \mhe~posterior distribution, which is more loosely constrained to be $0.56^{+0.33}_{0.37}$\msun.

%talk about 2007gr
Next, we present our fits to the \lc~of SN~2007gr in Figure~\ref{fig:07gr_lc_fits}. SN~2007gr is a Type Ic SN in the nearby galaxy NGC~1058 ($d_L\simeq9.3$\,Mpc) and is among the best-observed and most extensively modeled SESNe, noted for its unusually narrow P-Cygni features and low expansion velocities \citep{Valenti2008, Hunter2009, Mazzali2010}. Following \cite{Hunter2009}, we adopt a milky way extinction of $E(B-V)=0.062$~mag and host $E(B-V)=0.03$~mag in our fitting procedure. Literature values of \mni~have range$\sim0.07-0.1$~\msun~and \mej~$\sim2-3.5$~\msun~from its \lc~and spectra. \cite{Yoon2019} is largely unable to constrain radioactive mixing within the ejecta, given limited color observations; however, they note that the observations are consistent with moderate mixing.

We find a posterior distribution of $0.075^{+0.005}_{-0.006}$~\msun~for \mni, of $3.787^{+0.833}_{-1.157}$~\msun~for \mej, and a \gamnn~posterior range of $0.377^{+0.134}_{-0.096}$, indicating an ejecta profile with \NI~moderately mixed into the outer layers. We also find that the bulk of the ejecta (the middle $\sim$90\% of the mass) travels between $\sim10^{3.5}$ and $\sim10^{3.9}~\rm km~s^{-1}$, with the outermost layers exceeding $10^{4}~\rm km~s^{-1}$. As before, the \mop~is well constrained to a range of $3.14^{+0.84}_{-1.00}$\msun, whereas the posterior distribution of \mhe~reproduces the prior distribution (i.e. is totally unconstrained). All values are consistent with the literature, and the measurable mixing is consistent with the findings of \cite{Yoon2019}. Interestingly, in contrast to SN~1994I, \cite{afsariardchi2021} measured a tail \mni~$\simeq0.047$, but do \textit{not} require an additional heating source to account for the discrepancy between the peak and tail (i.e., their $\beta>0$).

%talk about iPTF13bvn
Finally, we discuss the \lc~of iPTF13bvn. iPTF13bvn is a nearby ($d_L\simeq22.5$\,Mpc) and the first SN~Ib with a progenitor detection \citep{Cao2013}. However, a binary progenitor became more persuasive in subsequent studies (see \citealt{Fremling2014} and \citealt{Srivastav2014} for a more detailed description). Its line-of-sight extinction remains uncertain: \cite{Cao2013} measure a host $E(B-V)=0.0437$~mag from the Na~{\sc i}~D lines (subsequently adopted by \citealt{Fremling2014} and \citealt{Reilly2016}), whereas \cite{Bersten2014} infer a higher $E(B-V)\simeq0.2$~mag from intrinsic colors (adopted by \citealt{Srivastav2014}). We adopt the lower \cite{Cao2013} value, the only direct sightline measurement and the one used by \cite{Fremling2014}, to whose parameters we compare. We adopt a foreground Milky Way extinction of $E(B-V)=0.0278$~mag \citep{Cao2013}. We infer \mni~of $0.05$~\msun, consistent with the \cite{Fremling2014} range $[0.037, 0.07]$~\msun, and an \mej~range of $[2.5,3.8]$~\msun, somewhat higher than their $[1.36,2.44]$~\msun. However, our fit finds $\sim1$\msun~of the fitted \mej~is \he, which is poorly constrained, whereas our fit finds an \mop~range of $[1.86, 2.49]$, which is potentially more consistent with \cite{Fremling2014}.

iPTF13bvn is the only Type~Ib among our three real \lcs, and its velocity profile is notably flatter and the \NI~is less mixed than those of the Type~Ic SNe~1994I and 2007gr. The middle $\sim$90\% of the ejecta mass sits near $\sim10^{3.5}~\rm km~s^{-1}$, rising to $\sim10^{4.5}~\rm km~s^{-1}$ in the fastest layers. Although Type~Ic SNe show systematically higher photospheric velocities than Type~Ib when averaged over their evolution \citep{Liu2016, Fremling2018}, whether the subtypes differ in the \emph{steepness} of their velocity profiles is little studied. \cite{Moriya2020} link weak \NI~mixing to a flatter early photospheric velocity evolution, qualitatively consistent with the flat, weakly mixed profile we infer. The degree of \NI~mixing in iPTF13bvn is itself contested: 1D \lc~models matching the fast rise require strong outward mixing \citep[to $\sim96\%$ of the ejecta mass;][]{Bersten2014, Fremling2014, Srivastav2014}, while the spectropolarimetry of \cite{Reilly2016} confines the \NI~within $\sim50$--$65\%$ of the ejecta radius and the color modeling of \cite{Yoon2019} favors weak mixing. Our fit supports the latter: we find \gamnn~of $[0.1, 0.2]$, weakly mixed and well below that of SN~1994I, in agreement with Y26 and \cite{Yoon2019}.

\section{Conclusions}
\label{sec:conclusion}

In this work, we present 1) a neural network emulator of the radiative transfer code \sed~and 2) a Bayesian inference pipeline using the emulator for physical parameter estimation from SESN \lcs. The emulator is a multilayer perceptron trained on a grid of $4499$ SESN \lcs~simulated in \sed~that span the nine-dimensional physical parameter space constructed by Y26. We construct this grid by augmenting the original Y26 grid of $1000$ \lcs~by a combination of random sampling and active learning. The emulator maps the nine Y26 physical parameters and time after explosion to a synthetic SESN spectrum. A single forward call to the emulator takes $\sim8$~milliseconds, which is a speedup of many orders of magnitude over direct \sed~simulation. This makes Bayesian inference over SESN \lcs~computationally feasible. We characterize two separate sources of systematic error in the inference pipeline: the error of the emulator in reproducing \sed~\lcs, and the error of \sed~in reproducing truly realistic SESN \lcs~as estimated by comparison to \cmf~\lcs~of identical models. Both sources of systematic error are incorporated directly into the \mos~fitting pipeline as additional uncertainties. We evaluate the inference pipeline on both simulated \lcs~designed to resemble those observed by ZTF and LSST. In all cases, the emulator-based inference recovers physical parameters more accurately than the classical Arnett model and produces reliable fits to the observed \lcs. The pipeline also reproduces three real SESN \lcs~well: SN~1994I, SN~2007gr, and iPTF13bvn, all of which were suggested by \cite{afsariardchi2021} as candidates for SESNe with power sources other than \NI~decay. Our key conclusions are:

\begin{enumerate}

    \item The emulator provides notably more accurate physical parameter recovery than the Arnett model. For simulated LSST-like \lcs, the emulator recovers \mni~with an RMS error of $0.008$~\msun, compared to $0.33$~\msun~for the Arnett model. Similarly, the emulator recovers \mej~with an RMS error of $0.9$~\msun, compared to $2.6$~\msun~for the Arnett model.

    \item Despite having a lower cadence in each filter than ZTF, LSST-like \lcs~yield similar recovery of \mej~and \vej~across the full parameter space explored here, while \NI~mixing is recovered comparably well for the two surveys. %The broad LSST filter set captures SESN color evolution well, making multi-band color evolution a particularly valuable diagnostic for physical inference.

    \item  The Arnett model's constraining capability lies in the degenerate combination $\rm \sqrt{m_{ej}/v_{ej}}$ rather than in constraining \mej~and \vej~separately because of the mass-velocity degeneracy that is implicit in the Arnett model. This degeneracy is weakened with our emulator. We find the joint posterior distributions of mass parameters and velocity parameters are weakly correlated for every fit we perform, suggesting that the emulator is able to largely independently constrain the ejecta mass and the ejecta velocity profile. 
    %The emulator recovers the combination  $\rm \sqrt{m_{ej}/v_{ej}}$ $\sim2$ times more accurately than the Arnett model, and recovers the individual parameters $\sim3$ times better. 
    Reproducing the true $\rm \sqrt{m_{ej}/v_{ej}}$ with the Arnett model further requires an effective gray opacity of $\kappa=0.054~\rm cm^2~g^{-1}$, below the commonly assumed $0.1-0.2~\rm cm^2~g^{-1}$.

    \item We apply our inference pipeline to three well-studied SESNe: SN~1994I, SN~2007gr, and iPTF13bvn. We find that our inferred \mni~for SN~2007gr and iPTF13bvn is, broadly, consistent with literature values. For SN~1994I our inferred \mni~($0.048$\msun) falls below the peak-based estimate of $\sim0.07$\msun~but matches the tail-based measurement of \cite{afsariardchi2021}. Our inferred \mej~is consistent with the literature for SN~2007gr, somewhat higher for iPTF13bvn, and roughly a factor of two above the literature value of $\sim1$\msun~for SN~1994I. We additionally constrain the degree of \NI~mixing in each event. The Type Ic SN~1994I and SN~2007gr are both moderately-to-strongly mixed, and the Type Ib iPTF13bvn is weakly mixed.

\end{enumerate}
In the next paper of this series, we plan to apply this inference pipeline on a statistical sample (of $\sim$hundreds) of SESNe. We aim to constrain the distribution of \NI~masses across the SESN population, explore systematic differences between Type Ib and Ic SN explosions, characterize the diversity of their ejecta profiles, and assess how often additional power sources are necessary to explain SESN \lcs. 

\bigskip
\section*{Acknowledgements}
The Villar Astro Time Lab acknowledges support through the David and Lucile Packard Foundation, the Research Corporation for Scientific Advancement (through a Cottrell Fellowship and Bridge Award) and the National Science Foundation under AST-2433718, AST-2407922 and AST-2406110. This work is supported
by the National Science Foundation under Cooperative Agreement PHY-2019786 (the NSF AI Institute for
Artificial Intelligence and Fundamental Interactions). 
M.R.D. acknowledges support from NSERC through grant RGPIN-2025-06224, the Canada Research Chairs Program (CRC-2023-00127), the Ontario ERA program (ER22-17-164) and the Dunlap Institute at
the University of Toronto. 
Analysis presented here includes code written using Claude and Chatgpt. 

\software{\texttt{numpy} \citep{numpy_cite}, \texttt{astropy}, \citep{astropy:2013, astropy:2018, astropy:2022}, \texttt{matplotlib} \citep{matplotlib}, \texttt{scipy} \citep{2020SciPy}, \sed~\citep{Kasen2006}, \texttt{pytorch}~\citep{pytorch}}

\bibliography{sample631}{}
\bibliographystyle{aasjournal}

\end{document}